\documentclass[12pt, draftclsnofoot, onecolumn]{IEEEtran}
\usepackage{graphicx}
\usepackage[english]{babel}
\usepackage{array}
\usepackage{colortbl}
\usepackage[english]{babel}
\usepackage[T1]{fontenc}
\usepackage{changepage}
\ifCLASSINFOpdf

\else

\fi
\usepackage{times}
\usepackage{amsmath}
\usepackage{supertabular}
\usepackage{cite}
\usepackage{color}
\usepackage{epsfig,graphics,subfigure,psfrag,amsmath,amssymb}

\usepackage{yfonts}

\usepackage{dsfont}

\usepackage{amsmath, amsthm, amssymb}
\usepackage{booktabs}
\usepackage{setspace}

\usepackage{stfloats}
\usepackage{amsfonts}
\usepackage{slashbox}
\usepackage{multirow}
\usepackage{makecell}
\usepackage{amssymb}
\usepackage{algorithmic}
\usepackage{hyperref}  
\hypersetup{
colorlinks=true,
linkcolor=black,
urlcolor=black,
citecolor=black,
}
\usepackage{url}
\usepackage{breakurl}
\usepackage{bm}
\usepackage{mathrsfs}
\usepackage{textcomp}

\usepackage[linesnumbered, ruled]{algorithm2e}

\usepackage{lineno}

\usepackage{changes}

\usepackage{cuted}
\usepackage{fancyhdr}

\begin{document}

\title{Beamforming Design and Trajectory Optimization for UAV-Empowered Adaptable Integrated Sensing and Communication}
\author{Cailian~Deng,
        Xuming~Fang,~\IEEEmembership{Senior~Member,~IEEE,}
        Xianbin Wang,~\IEEEmembership{Fellow,~IEEE}
\thanks{C. Deng and X. Fang are with the Key Laboratory of Information Coding and Transmission, Southwest Jiaotong University, Chengdu 611756, China. E-mail: dengcailian@my.swjtu.edu.cn; xmfang@swjtu.edu.cn. (Corresponding author: Xuming Fang.)}
\thanks{X. Wang is with the Department of Electrical and Computer Engineering, Western University, London, ON N6A 5B9, Canada. E-mail: xianbin.wang@uwo.ca.}
}

\maketitle
\thispagestyle{fancy}

\lfoot{~\copyright~2022 IEEE. This work has been submitted to the IEEE for possible publication. Copyright may be transferred without notice, after which this version may no longer be accessible.}
%

\maketitle

\begin{abstract}
Unmanned aerial vehicle (UAV) has high flexibility and controllable mobility, therefore it is considered as a promising enabler for future integrated sensing and communication (ISAC). In this paper, we propose a novel adaptable ISAC (AISAC) mechanism in the UAV-enabled system, where the UAV performs sensing on demand during communication and the sensing duration is configured flexibly according to the application requirements rather than keeping the same with the communication duration. Our designed mechanism avoids the excessive sensing and waste of radio resources, therefore improving the resource utilization and system performance. In the UAV-enabled AISAC system, we aim at maximizing the average system throughput by optimizing the communication and sensing beamforming as well as UAV trajectory while guaranteeing the quality-of-service requirements of communication and sensing. To efficiently solve the considered non-convex optimization problem, we first propose an efficient alternating optimization algorithm to optimize the communication and sensing beamforming for a given UAV location, and then develop a low-complexity joint beamforming and UAV trajectory optimization algorithm that sequentially searches the optimal UAV location until reaching the final location. Numerical results validate the superiority of the proposed adaptable mechanism and the effectiveness of the designed algorithm.
\end{abstract}


\IEEEpeerreviewmaketitle

\section{Introduction}\label{section:1}
\IEEEPARstart{I}{n} recent years, there have been growing concurrent communication and sensing demands on emerging platforms such as autonomous-driving, unmanned aerial vehicles (UAVs), and remote healthcare, which leads to a new paradigm referred to as integrated sensing and communication (ISAC) \cite{8972666Survey, 8999605_MIMO, 9585321Survey}. By sharing the same spectrum resources and wireless infrastructures between sensing and communication systems, ISAC can provide a number of advantages over the traditional fully separated stand-alone systems, such as significantly reduced cost, and enhanced spectrum efficiency, energy efficiency, and hardware utilization efficiency.

Beamforming plays an important role in improving the system performance of ISAC. Recently, numerous beamforming design approaches have been proposed to provide high-quality ISAC services in various ISAC systems \cite{9124713_MIMO,9724174GeneralizedBeamforming,8288677_MIMO, horizontal_locations,fang2022joint}. For instance, in the ISAC system where sensing works with the downlink (DL) communication simultaneously \cite{9124713_MIMO}, a transmit beamforming scheme was proposed to maximize the sensing performance while guaranteeing the signal-to-interference-plus-noise ratio (SINR) requirement of each communication user. In \cite{8288677_MIMO}, effective transmit beamforming schemes were developed for the separated sensing and communication antenna deployment as well as the shared antenna deployment in the ISAC system. By carefully designing the transmit beamforming for different antenna deployments, the ideal sensing beampattern can be approached under the power budget and communication SINR constraints. Receiver technology is also a factor that affects the transmit beamforming design. In \cite{horizontal_locations}, the authors considered two types of communication receivers with and without sensing interference cancellation capabilities, and designed the optimal beamforming for the two types of receivers separately. Besides, to further explore the potential of the ISAC system, some works have started investigating the emerging technologies and approaches to enhance the overall system performance of ISAC, such as reconfigurable intelligent surface-assisted ISAC \cite{9416177RIS, 9769997RIS, 9591331RIS} and artificial intelligence-enabled ISAC \cite{9449980AI, 9492131AI}.

However, the related prior works on ISAC mainly focus on terrestrial networks, which have many inherent limitations, such as signal blockage caused by the surrounding obstacles and scatters on the ground, imperfect coverage due to the limited available infrastructures, and vulnerability to natural disasters or artificial infrastructure damages, resulting in serious performance degradation or even unavailability of ISAC service. Fortunately, UAV has been envisioned as a cost-effective aerial platform to overcome the above limitations and enable ISAC on demand since it has the high mobility, flexibility and controllability and can provide the strong air-ground line-of-sight (LoS) channels \cite{7888557_EnergyEfficientUAV, 7572068RelayUAV, 8618602SecuringUAV, 8247211MUltiUAV,8647530UAVSensing, 9409835UAVSensing, 8894454UAVSensing}. In the UAV-enabled ISAC system, in order to fully exploit the design degrees of freedom for improving the system performance, we also need to carefully investigate the UAV location or trajectory optimization in addition to designing resource allocation and beamforming. How to jointly design the resource allocation, beamforming, and UAV location or trajectory for unleashing the maximum potential of the UAV-enabled ISAC system is an extremely appealing yet challenging problem. At present, the research on the UAV-enabled ISAC system is still in its infancy, and only a few works focused on solving the above joint design problem. In \cite{lyu2021joint}, the UAV trajectory and beamforming were jointly designed to improve the system performance of the UAV-enabled ISAC system. To further enhance the communication and sensing coverage and increase the integration gain of communication and sensing, a multi-UAV cooperative ISAC system was proposed in \cite{9293257Constrained}, in which the user association, UAV location, and power allocation were jointly optimized to maximize the total network utility while guaranteeing the sensing accuracy.

In the above works on UAV-enabled ISAC systems, sensing and communication always work simultaneously and have the same duration. In such a configuration, sensing is performed with fixed time and the sensing duration remains unchanged, which ignores the fact that different sensing applications usually have different requirements on the sensing duration. The sensing duration mainly consists of two basic components, i.e., the sensing dwell duration and sensing interval \cite{skolnik2008radar}, where the sensing dwell duration is the time over which a position is illuminated and the sensing interval is the time between two repetitions of sensing. While the sensing dwell duration and sensing interval is typically set according to the sensing application type (e.g., scanning, detection, or tracking) and the desired sensing quality (e.g., the probability of false alarms or sensing accuracy). In the modern sensing/radar systems, multi-pulse accumulation technology is widely used for improving the signal-to-noise ratio (SNR) of the target reflected echo \cite{skolnik2008radar}. The length of the dwell duration has a direct effect on the echo SNR, thereby the sensing performance can be adjusted by controlling the dwell duration. For example, we can configure a long dwell duration when scanning a specified area to maximize the cumulative detection range or minimize the probability of false alarms. Generally, the sensing interval depends on the target kinematics and the desired tracking continuity and accuracy. In the scenario with high-speed moving targets, a relatively short sensing interval is required to prevent losing targets or large location error in tracking, while for the scenario with low-speed moving targets, too frequent sensing is unnecessary, and the sensing interval can be set relatively long to avoid waste of radio resources. Ignoring the practical sensing duration requirement and enabling simultaneous sensing and communication all the time during the ISAC service period may cause excessive sensing and waste of spectrum and energy, which may seriously degrade the overall system performance. Therefore, it is necessary to reasonably set the sensing duration according to the practical application requirements for unleashing the maximum potential of ISAC, especially in the resource-limited UAV-enabled networks.

Motivated by the above considerations, we study a UAV-enabled ISAC system that simultaneously provides communication services for ground multiple users and sensing services during the ISAC service period. Unlike the previous works that sensing and communication keep the same duration, we propose an adaptable ISAC (AISAC) mechanism that can flexibly configure the sensing duration according to the practical application requirements and enable sensing on demand during the communication rather than enabling sensing all the time, thus offering additional opportunities to improve the resource utilization and system performance. Our objective is to maximize the average system throughput by optimizing the UAV trajectory as well as the communication and sensing beamforming while guaranteeing the quality-of-service (QoS) requirements of communication and sensing. To efficiently solve this problem, we propose a low-complexity joint beamforming and UAV trajectory optimization algorithm to find a high-quality solution. The main contributions of this paper are summarized as follows:
\begin{itemize}
  \item We consider a UAV-enabled ISAC system, where the UAV communicates with ground multiple users and simultaneously performs sensing during the ISAC service period. To cater for different sensing duration requirements of different ISAC services and avoid potential resource waste and system performance degradation, we propose an AISAC mechanism that flexibly configures the sensing duration according to the practical application requirements and enables sensing on demand during the communication rather than enabling sensing all the time.
  \item Based on the proposed AISAC mechanism, we formulate an optimization problem to maximize the average system throughput by optimizing the communication and sensing beamforming as well as UAV trajectory while guaranteeing the QoS requirements of communication and sensing. To efficiently solve the problem, we first propose an efficient  alternating optimization algorithm to optimize the communication and sensing beamforming for a given UAV location, and then develop a low-complexity joint beamforming and UAV trajectory optimization algorithm that sequentially searches the optimal UAV location until reaching the final UAV location.
  \item By simulation, we find the proposed AISAC mechanism can significantly improve the average system throughput by flexibly configuring the sensing duration and enabling sensing on demand. Furthermore, we find that the proposed joint beamforming and UAV trajectory optimization algorithm can achieve a significant improvement on the average system throughput compared to other benchmark schemes.
\end{itemize}

The rest of the paper is organized as follows. In Section \ref{section:2}, we introduce the system model and designed AISAC mechanism, and formulate the average system throughput maximization problem. In Section \ref{section:3}, we first propose an efficient optimization algorithm to optimize the communication and sensing beamforming for a given UAV location, and then develop a low-complexity joint beamforming and UAV trajectory optimization algorithm to solve the formulated maximization problem. In Section \ref{section:4}, we present the numerical results to validate the performance of the proposed algorithm. Finally, we conclude this work in Section \ref{section:5}.

\emph{Notations}: Boldface letters refer to vectors (lower case) or matrices (upper case). $\mathbf{A}\succeq \mathbf{0}$ means that $\mathbf{A}$ is positive
semidefinite. $\mathbf{0}$ denotes an all-zero vector or matrix. $\mathbf{1}$ denotes an all-one vector. $\mathbb{I}$ denotes an identity matrix. $\mathrm{rank}(\mathbf{A})$ and $\mathrm{tr}(\mathbf{A})$ denote the rank and trace of matrix $\mathbf{A}$, respectively. The superscript $(\cdot)^{H}$ denote the conjugate transpose operator. $|.|$ and $\|.\|$ denote the magnitude of a complex number and Euclidean norm, respectively. $\mathbb{C}$ denotes the complex space. Random variable $x\sim \mathcal{CN}(\mu,\sigma^{2})$ follows the distribution of a complex Gaussian with mean $\mu$ and variance $\sigma^{2}$, and random vector $\mathbf{x}\sim \mathcal{CN}(\bm{\mu},\bm{\sum})$ follows the distribution of a complex Gaussian with mean vector $\bm{\mu}$ and covariance matrix $\bm{\sum}$.

\section{System Model and Problem Formulation}\label{section:2}
\begin{figure}[!t]
\renewcommand{\figurename}{Fig.}
  \centering
  \includegraphics[width=9cm]{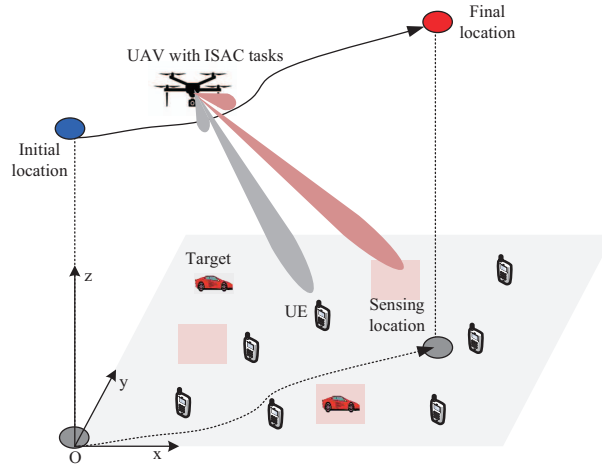}
  \caption{The proposed UAV-enabled AISAC system. The UAV communicates with multiple users and simultaneously enables sensing on demand during the ISAC service period.}
\label{Fig:1a}
\end{figure}

We consider a UAV-enabled AISAC system as shown in Fig. \ref{Fig:1a}, in which the UAV flies from the predetermined initial location to the final location within an ISAC service period $T$. At the end of period $T$, the UAV needs to arrive at the final location (e.g., launch/landing sites) for some practical reasons, such as recharging or refueling. During the period $T$, the UAV equipped with a vertically placed $Q$-antenna uniform linear array (ULA) communicates with $K$ single-antenna ground user equipments (UEs) and simultaneously senses potential targets on demand at the interested ground area. The UAV performs sensing using its transmit signals, i.e., the UAV receives the echo signals that are transmitted by itself. Similar to the prior works \cite{9124713_MIMO}, \cite{horizontal_locations}, we suppose that the UAV senses potential targets at the finite $J$ locations that are predetermined by specific sensing tasks. For example, if the UAV performs the detection task without knowing the presence and exact locations of potential targets, the sensing locations can be set uniformly distributed over the whole interested area. By contrast, if the UAV performs the target tracking and roughly knows a-priori locations of potential targets, then the sensing locations can be set as the possible locations of targets.

\begin{figure}[!t]
\renewcommand{\figurename}{Fig.}
  \centering
  \includegraphics[width=9cm]{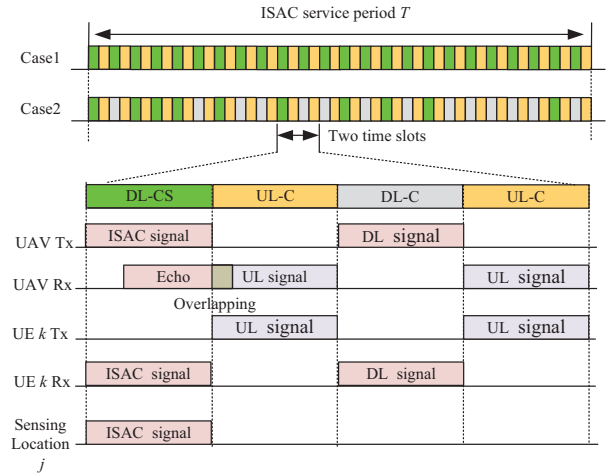}
  \caption{AISAC time slot structure. DL-CS, DL-C, and UL-C indicate DL communication and sensing, DL communication, and UL communication, respectively. Case1: The UAV performs communication and sensing in every time slot. Case 2: The UAV performs sensing only in part of time slots.}
\label{Fig:1b}
\end{figure}

\subsection{AISAC Time Slot Structure}
For convenience, we divide the period $T$ into $N$ time slots with a slot length $\tau= T/N$, which is indexed by $n\in \{1,2,\ldots,N\}$. Each time slot operates in a time division duplex (TDD) model and consists of the DL duration and UL duration\footnote{Note that the guard period is typically required and used in TDD protocols. Guard period can be set very small compared to the DL duration and UL duration and is negligible for simplicity.}, as illustrated in Fig. \ref{Fig:1b}. In each time slot, the UAV performs communication procedures while adaptively enabling sensing according to the practical sensing requirements. For simplicity, we classify the services into two types: communication only and ISAC. Specifically, in the former, the UAV transmits the DL communication signals during the DL stage and receives the UL communication signals from UEs during the UL stage. In the latter, the UAV transmits the ISAC signals to enable DL communication and sensing and may receive the echos reflected from the sensing locations during the DL stage, and receives sensing echos reflected from the ground mixed with the UL communication signals during the UL stage. The communication and sensing share the same antenna array at the UAV, and the ISAC signals are a superimposition of individually precoded communication and sensing signals.

\subsection{AISAC Mechanism}
In practical sensing applications, too long sensing duration may result in unnecessary resource usage, while too short sensing duration may lead to the poor sensing performance. Therefore, it is necessary to develop an effective way to reasonably configure the sensing duration for improving the resource utilization and system performance. We define $\psi_{n}\in\{0,1\}$ as the sensing indicator at time slot $n$, where $\psi_{n}=1$ indicates that the UAV performs sensing during the DL communication to enable ISAC while $\psi_{n}=0$ means that the UAV only enables DL communication. Therefore, the sensing duration can be configured by setting the value of $\psi_{n}$. For instance, if a longer sensing duration is intended, the sensing indicator of more time slots should be set to $1$.

In the following, we propose a novel AISAC mechanism that determines $\psi_{n}$ according to the variations of sensing-related parameters. Specifically, we use a ratio $f_{m}^{\mathrm{pv}}$ to characterize the parameter difference between the $(m-1)$-th sensing and $m$-th sensing, i.e.,
\begin{equation}
\begin{aligned}
f_{m}^{\mathrm{pv}}=\frac{|\zeta_{m}-\zeta_{m-1}|}{\zeta_{m-1}},
\label{equa_0}
\end{aligned}
\end{equation}
where $\zeta_{m}$ denotes the sensing-related parameter of the $m$-th sensing, such as the target location, target moving speed, or error/mean squared error (MSE) of the target location/velocity estimation \cite{8894454UAVSensing,9293257Constrained}. $f_{m}^{\mathrm{pv}}$ reflects the demand level of the sensing duration, thus can be applied to guide the setting of $\psi_{n}$. The smaller/larger the $f_{m}^{\mathrm{pv}}$, the shorter/longer the required sensing duration, and the larger/smaller the sensing interval, where the sensing interval is the time between two adjacent sensing. By setting a threshold for $f_{m}^{\mathrm{pv}}$, i.e., $f_{\mathrm{th}}^{\mathrm{pv}}$, the sensing interval can be configured. Specifically, when $f_{m}^{\mathrm{pv}}$ is smaller/greater than a given threshold, we can increase/decrease, such as in exponentially, the sensing interval, thereby allowing higher sensing flexibility compared to the fixed sensing interval. Considering that the ever increasing of the sensing interval may lead to the loss of tracked targets or missing detection of new targets, we can set a maximum value, denoted by $\Delta_{\mathrm{max}}$, for the sensing interval instead of increasing all the time. Similarly, we can also set a minimum value, denoted by $\Delta_{\mathrm{min}}$, to prevent the ever decreasing of the sensing interval. The proposed AISAC mechanism is summarized in Algorithm \ref{alg:Iterative_0}.

\begin{algorithm} [!t]
\setlength\abovedisplayskip{1pt}
\setlength\belowdisplayskip{1pt}
	\caption{Adaptable ISAC Mechanism.}\label{alg:Iterative_0}
	\KwIn{$m$.}

\Repeat{The configuration of the sensing interval within the period $T$ is obtained}{
Compute $f_{m}^{\mathrm{pv}}=\frac{|\zeta_{m}-\zeta_{m-1}|}{\zeta_{m-1}}$;\\
\uIf{$f_{m}^{\mathrm{pv}}\leq f_{\mathrm{th}}^{\mathrm{pv}}$}{
Increase the sensing interval to $\tilde{\Delta}_{m}$, and then the implemented sensing interval is $\Delta_{m}=\mathrm{min}\{\tilde{\Delta}_{m},\Delta_{\mathrm{max}}\}$;\\} \Else{Decrease the sensing interval to $\tilde{\Delta}_{m}$, and then the implemented sensing interval is $\Delta_{m}=\mathrm{max}\{\tilde{\Delta}_{m},\Delta_{\mathrm{min}}\}$;\\}
$m=m+1$;
}
		\KwOut{The configuration of the sensing interval.}
\end{algorithm}

\subsection{Transmit Signal Model}
\subsubsection{UAV Transmit Signal Model}
During the DL stage, the transmit signal of the UAV at time slot $n$ is expressed as $\mathbf{x}_{n}^{\mathrm{DL}}=\sum_{k=1}^{K}\mathbf{w}_{k,n}s_{k,n}^{\mathrm{DL}}
+\psi_{n}\sum_{j=1}^{J}\mathbf{r}_{j,n}s_{j,n}^{\mathrm{Sens}}$,
where $\mathbf{w}_{k,n}\in \mathbb{C}^{Q\times 1}$ and $\mathbf{r}_{j,n}\in \mathbb{C}^{Q\times 1}$ are the transmit beamforming vector for DL communication and sensing, respectively. $s_{k,n}^{\mathrm{DL}}\in \mathbb{C}$ and $s_{j,n}^{\mathrm{Sens}}\in \mathbb{C}$ are the communication signal associated with UE $k$ and dedicated sensing signal associated with the sensing location $j$, respectively, which are uncorrelated and independent random variables with zero mean and unit variance.

\subsubsection{UE Transmit Signal Model}
During the UL stage, the transmit signal of UE $k$ at time slot $n$ is $x_{k,n}^{\mathrm{UL}}=\sqrt{p_{k}}s_{k,n}^{\mathrm{UL}}$, where $p_{k}$ denotes the transmit power. $s_{k,n}^{\mathrm{UL}}\in \mathbb{C}$ is the communication signal, which is independent random variable with zero mean and unit variance.

\subsection{Receive Signal Model}

\subsubsection{Communication Receive Signal Model}
We consider a three-dimensional (3D) Cartesian coordinate system, in which the horizontal coordinate of UE $k$ is $\mathbf{q}_{k}^{\mathrm{UE}}=(x_{k}, y_{k})$. For convenience, we consider that the UAV trajectory within the period $T$ is discretized into $(N-1)$ line segments which can be represented by $N$ horizontal locations/waypoints $\{\mathbf{q}_{n}=(x_{n}, y_{n}),\forall n\}$. Time slot length is sufficiently small so that the UAV flies through a small distance in a time slot and the horizontal location of the UAV is approximately unchanged within a time slot. Due to the relatively high altitude of the UAV, there generally exists the strong LoS links between the UAV and UEs. Thus, the wireless channel between the UAV and UE $k$ at time slot $n$ can be modeled as
\begin{equation}
\mathbf{h}_{k,n}^{\mathrm{DL}} = \mathbf{h}_{k,n}^{\mathrm{UL}}=\sqrt{\frac{\beta_{0}}{d_{k,n}^{2}}}\mathbf{a}(\theta_{k,n}),\label{channel_1}
\end{equation}
where the reciprocity of the DL and UL channels are assumed, $\beta_{0}$ denotes the channel power gain at a reference distance $d_{0}=1$ m, $d_{k,n} = \sqrt{H_{c}^{2}+\|\mathbf{q}_{n}-\mathbf{q}_{k}^{\mathrm{UE}}\|^{2}}$ is the distance between the UAV and UE $k$, and $\mathbf{a}(\theta_{k,n}) = [1,e^{-j\pi\cos\theta_{k,n}},...,e^{-j\pi(Q-1)\cos\theta_{k,n}}]^{T}$ denotes the transmit steering vector towards UE $k$. $H_{c}$ is the flying height of the UAV, and $\theta_{k,n}=\arccos \big(H_{c}/\sqrt{H_{c}^{2}+\|\mathbf{q}_{n}-\mathbf{q}_{k}^{\mathrm{UE}}\|^{2}}\big)$ is the angle of departure (AoD) corresponding to UE $k$.

At time slot $n$, the signal received by UE $k$ and the combined signal at the UAV can be written as
\begin{align}
&y_{k,n}^{\mathrm{DL}}=\underbrace{(\mathbf{h}_{k,n}^{\mathrm{DL}})^{H}\mathbf{w}_{k,n}s_{k,n}^{\mathrm{DL}}}_{\text{Intended signal}}+\underbrace{\sum_{i\neq k}^{K}(\mathbf{h}_{k,n}^{\mathrm{DL}})^{H}\mathbf{w}_{i,n}s_{i,n}^{\mathrm{DL}}}_{\text{Inter-user interference}}
+\psi_{n}\underbrace{\sum_{j=1}^{J}(\mathbf{h}_{k,n}^{\mathrm{DL}})^{H}\mathbf{r}_{j,n}}_{\text{Sensing interference}}
+\underbrace{n_{k}^{\mathrm{DL}}}_{\text{AWGN noise}},\label{Signal_DL}\\
&y_{k,n}^{\mathrm{UL}}=\underbrace{\mathbf{v}_{k,n}^{H}\mathbf{h}_{k,n}^{\mathrm{UL}}x_{k,n}^{\mathrm{UL}}}
_{\text{Intended signal}}+
\underbrace{\mathbf{v}_{k,n}^{H}\sum_{i\neq k}^{K}\mathbf{h}_{i,n}^{\mathrm{UL}}x_{i,n}^{\mathrm{UL}}}_{\text{Inter-user interference}}
+\underbrace{\mathbf{v}_{k,n}^{H}
\bigg(\xi_{n}\sum_{j=1}^{J}\mathbf{G}_{j,n}\mathbf{x}_{n}^{\mathrm{DL}}\bigg)}
_{\text{Sensing interference}}+ \underbrace{\mathbf{v}_{k,n}^{H}\mathbf{n}_{k}^{\mathrm{UL}}}_{\text{AWGN noise}},\label{Signal_UL}
\end{align}
where $\mathbf{v}_{k,n}\in \mathbb{C}^{Q\times 1}$ is the unit-norm receive beamforming vector. $n_{k}^{\mathrm{DL}}\sim \mathcal{CN}(0,\sigma_{\mathrm{DL}}^{2})$ and $\mathbf{n}_{k}^{\mathrm{UL}}\sim \mathcal{CN}(\mathbf{0},\sigma_{\mathrm{UL}}^{2}\mathbb{I}_{Q})$ are the complex additive white Gaussian noise (AWGN) at UE $k$ and at the UAV, respectively. $\xi_{n}\in\{0,1\}$ is the signal overlapping indicator, where $\xi_{n}=1$ indicates that the UL communication signals are overlapped with the echos, otherwise $\xi_{n}=0$. Note that $\xi_{n}=0$ holds if $\psi_{n}=0$. It is assumed that the signal components originating from DL communication and other background clutter have been eliminated by proper signal processing, therefore, equation (\ref{Signal_UL}) only includes UL communication signals and/or the signals reflected from $J$ sensing locations.

The SINR of DL and UL communication for UE $k$ are calculated by
\begin{equation}
\begin{aligned}
\gamma_{k,n}^{\mathrm{DL}} &=\frac{\big|(\mathbf{h}_{k,n}^{\mathrm{DL}})^{H}\mathbf{w}_{k,n}\big|^{2}}{\mathbf{I}_{k,n}^{\mathrm{DL}}
+\sigma_{\mathrm{DL}}^{2}},
\gamma_{k,n}^{\mathrm{UL}}
&=\frac{p_{k}\big|\mathbf{v}_{k,n}^{H}\mathbf{h}_{k,n}^{\mathrm{UL}}\big|^{2}}
{\mathbf{I}_{k,n}^{\mathrm{UL}}+\sigma_{\mathrm{UL}}^{2}},
\end{aligned}
\end{equation}
where $\mathbf{I}_{k,n}^{\mathrm{DL}} = \big|\sum_{i\neq k}^{K}(\mathbf{h}_{k,n}^{\mathrm{DL}})^{H}\mathbf{w}_{i,n}
+\psi_{n}\sum_{j=1}^{J}(\mathbf{h}_{k,n}^{\mathrm{DL}})^{H}\mathbf{r}_{j,n}\big|^{2}$ and $\mathbf{I}_{k,n}^{\mathrm{UL}}=\big|\mathbf{v}_{k,n}^{H}\big(\sum_{i\neq k}^{K}\sqrt{p_{i}}\mathbf{h}_{i,n}^{\mathrm{UL}}
+\xi_{n}\sum_{j=1}^{J}\mathbf{G}_{j,n}\mathbf{x}_{n}^{\mathrm{DL}}\big)\big|^{2}$ represent the co-channel interference for DL and UL, respectively. Then, the DL and UL throughput in bit/s/Hz for UE $k$ at time slot $n$ can be computed by
\begin{equation}
\begin{aligned}
R_{k,n}^{\mathrm{DL}}&=\frac{t_{n}^{\mathrm{DL}}}{\tau/2}\log_{2}(1+\gamma_{k,n}^{\mathrm{DL}}),
R_{k,n}^{\mathrm{UL}}&=\frac{t_{n}^{\mathrm{UL,1}}}{\tau/2}\log_{2}\big(1+\gamma_{k,n}^{\mathrm{UL}}|_{\xi_{n}=1}\big)
+\frac{t_{n}^{\mathrm{UL,2}}}{\tau/2}\log_{2}\big(1+\gamma_{k,n}^{\mathrm{UL}}|_{\xi_{n}=0}\big),
\end{aligned}
\end{equation}
where $t_{n}^{\mathrm{DL}}$, $t_{n}^{\mathrm{UL,1}}$, and $t_{n}^{\mathrm{UL,2}}$ denote the DL duration, UL duration when $\xi_{n}=1$, and UL duration when $\xi_{n}=0$, respectively. Note that when no echo interference in time slot $n$, i.e., $\psi_{n} = 0$, $t_{n}^{\mathrm{UL,1}}=0$ and  $t_{n}^{\mathrm{UL,1}}\log_{2}\big(1+\gamma_{k,n}^{\mathrm{UL}}|_{\xi_{n}=1}\big)=0$ hold. $\gamma_{k,n}^{\mathrm{UL}}|_{\xi_{n}=1}$ and $\gamma_{k,n}^{\mathrm{UL}}|_{\xi_{n}=0}$ denote the SINR conditioned on $\xi_{n}=1$ and $\xi_{n}=0$, respectively.

\subsubsection{Sensing Receive Signal Model}
The signals transmitted by the UAV are reflected from the sensing locations and then received by the UAV, therefore the sensing channel corresponding to the sensing location $j$ can be modeled as
\begin{equation}
\mathbf{G}_{j,n}=\alpha_{j,n}\mathbf{b}(\theta_{j,n})\mathbf{b}(\theta_{j,n})^{H}e^{-j2\pi f_{c}\tau_{j,n}},\label{channel_2}
\end{equation}
where $f_{c}$ and $\tau_{j,n} = 2d_{j,n}/c$ represent the carrier frequency and round-trip delay, respectively. $\alpha_{j,n} = \sigma_{j}/(2d _{j,n})$ is the reflection coefficient, where $\sigma_{j}$ represents the complex radar cross-section (RCS) \cite{9557830YuanWeijie_twoway}, \cite{9171304LiuFan_twoway}, $d_{j,n}=\sqrt{H_{c}^{2}+\|\mathbf{q}_{n}-\mathbf{q}_{j,n}^{\mathrm{Sens}}\|^{2}}$ is the distance between the UAV and sensing location $j$, and $\mathbf{q}_{j,n}^{\mathrm{Sens}}=(x_{j,n},y_{j,n})$ is the coordinate of the sensing location $j$. $\mathbf{b}(\theta_{j,n})=[1,e^{-j\pi\cos\theta_{j,n}},..., e^{-j\pi(Q-1)\cos\theta_{j,n}}]^{T}$
denotes the transmit steering vector, where $\theta_{j,n}$ is the AoD corresponding to the sensing location $j$ and given as $\theta_{j,n}=\arccos\big(H_{c}/\sqrt{H_{c}^{2}+\|\mathbf{q}_{n}-\mathbf{q}_{j,n}^{\mathrm{Sens}}\|^{2}}\big)$.

The communication signals interfere the reflected echoes, which degrades the sensing performance. To improve the sensing performance, we propose that the UAV extracts sensing signals by subtracting the known DL communication signals and decoded UEs' signals from the received signals. Besides, the DL communication signals reflected by the sensing locations are also utilized for sensing in our considered system. Note that the self-interference caused by simultaneous transmission and reception during sensing is assumed to be cancelled by employing proper self-interference cancellation methods.

The combined sensing signal at the UAV is expressed as
\begin{align}
y_{j,n}^{\mathrm{Sens}}&=\underbrace{\mathbf{u}_{j,n}^{H}\mathbf{G}_{j,n}
\big(\mathbf{r}_{j,n}+\sum_{k=1}^{K}\mathbf{w}_{k,n}s_{k,n}^{\mathrm{DL}}\big)}_{\text{Intended sensing signal}}
+\underbrace{\mathbf{u}_{j,n}^{H}\sum_{i\neq j}^{J}\mathbf{G}_{i,n}\big(\mathbf{r}_{i,n}+\sum_{k=1}^{K}\mathbf{w}_{k,n}s_{k,n}^{\mathrm{DL}}\big) }_{\text{Clutter interference}} +\underbrace{\mathbf{u}_{j,n}^{H}\mathbf{n}_{j}^{\mathrm{Sens}}}_{\text{AWGN noise}},\label{Signal_Sens}
\end{align}
where $\mathbf{u}_{j,n}\in\mathbb{C}^{Q\times 1}$ is the unit-norm receive beamforming vector corresponding to the sensing location $j$. $\mathbf{n}_{j}^{\mathrm{Sens}}\sim \mathcal{CN}(\mathbf{0},\sigma_{\mathrm{Sens}}^{2}\mathbb{\mathbf{I}}_{Q})$ is the AWGN at the UAV. Assuming that all the reflected echoes at time slot $n$ are combined for sensing processing \cite{9226446Gain}, then the SINR corresponding to the sensing location $j$ is given by
\begin{equation}
\begin{aligned}
\gamma_{j,n}^{\mathrm{Sens}}&=\frac{G_{p}\big|\mathbf{u}_{j,n}^{H}\mathbf{G}_{j,n}
\big(\mathbf{r}_{j,n}+\sum_{k=1}^{K}\mathbf{w}_{k,n}\big)\big|^{2}}
{\mathbf{I}_{j,n}^{\mathrm{Sens}}+\sigma_{\mathrm{Sens}}^{2}},
\end{aligned}
\end{equation}
where $\mathbf{I}_{j,n}^{\mathrm{Sens}} = \big|\mathbf{u}_{j,n}^{H}\sum_{i\neq j}^{J}\mathbf{G}_{i,n}\big(\mathbf{r}_{i,n}+\sum_{k=1}^{K}\mathbf{w}_{k,n}\big)\big|^{2}$ represents the co-channel interference and $G_{p}$ is the processing gain brought by the sensing processing operation\footnote{Note that the detailed sensing data processing and analysis are neglected in this work and designated to our future work.} \cite{9226446Gain}.

\subsection{Problem Formulation}
In the UAV-enabled AISAC system, we define the average system throughput as the sum of average DL and UL throughput over $N$ time slots, i.e., $\frac{1}{N}\sum_{n=1}^{N}\sum_{k=1}^{K}\big( R_{k,n}^{\mathrm{DL}}+ R_{k,n}^{\mathrm{UL}}\big)$. We aim to maximize the average system throughput by jointly optimizing the communication and sensing beamforming $\{\mathbf{w}_{k,n},\mathbf{v}_{k,n},\mathbf{r}_{j,n},\mathbf{u}_{j,n},\forall k,j,n\}$ as well as UAV trajectory $\{\mathbf{q}_{n},\forall n\}$, subject to the QoS requirements of communication and sensing, UAV's transmit power constraints, and UAV location constraints. The optimization problem is formulated as
\begin{subequations} \label{P}
\begin{align}
&{P_{0}:}\underset{\{\mathbf{w}_{k,n},\mathbf{v}_{k,n},\mathbf{r}_{j,n},
\mathbf{u}_{j,n},\mathbf{q}_{n},\forall k,j,n\}}
{\mathrm{max}}\begin{aligned}&\frac{1}{N}\sum_{n=1}^{N}\sum_{k=1}^{K}\big( R_{k,n}^{\mathrm{DL}}+ R_{k,n}^{\mathrm{UL}}\big)\end{aligned} \notag
\\& \quad \mathrm{s.t.} \quad \psi_{n}\gamma_{j,n}^{\mathrm{Sens}}\geq \psi_{n}\gamma_{\mathrm{th}}^{\mathrm{Sens}},\forall j,n,\label{Pa}
\\& \quad\quad\quad  \gamma_{k,n}^{\mathrm{DL}}\geq\gamma_{\mathrm{th}}^{\mathrm{DL}},\forall k,n,\label{Pb}
\\& \quad\quad\quad  \gamma_{k,n}^{\mathrm{UL}}\geq\gamma_{\mathrm{th}}^{\mathrm{UL}},\forall k,n,\label{Pc}
\\& \quad\quad\quad \sum_{k=1}^{K}\big\|\mathbf{w}_{k,n}\big\|^{2}
+\psi_{n}\sum_{j=1}^{J}\big\|\mathbf{r}_{j,n}\big\|^{2}\leq P_{\mathrm{max}}^{\mathrm{uav}},\forall n, \label{Pd}
\\& \quad\quad\quad \big\|\mathbf{v}_{k,n}\big\|^{2}=1,
\psi_{n}\big\|\mathbf{u}_{j,n}\big\|^{2}=\psi_{n}, \forall k,j,n,\label{Pe}
\\& \quad\quad\quad \|\mathbf{q}_{n+1}-\mathbf{q}_{n}\|^{2} \leq V_{\mathrm{max}}\tau, \forall n, \label{Pf}
\\& \quad\quad\quad \mathbf{q}_{1}=\mathbf{q}_{I}, \mathbf{q}_{N}=\mathbf{q}_{F}, \label{Pg}
\end{align}
\end{subequations}
where $\gamma_{\mathrm{th}}^{\mathrm{Sens}}$ is the minimum SINR requirement of sensing. Constraint (\ref{Pa}) guarantees the success of sensing at time slot $n$. Constraints (\ref{Pb}) and (\ref{Pc}) describe the QoS requirements of DL and UL communication. Constraint (\ref{Pd}) is the power limit of the UAV, where $P_{\mathrm{uav}}^{\mathrm{max}}$ is the maximum transmit power at the UAV. Constraints (\ref{Pe}) denote the unit norm constraints of the receive beamforming vectors. Constraint (\ref{Pf}) means that the flying speed of the UAV cannot exceed its maximum value $V_{\mathrm{max}}$. $\mathbf{q}_{I}$ and $\mathbf{q}_{F}$ in constraint (\ref{Pg}) denote the initial and final horizontal locations of the UAV.

\section{Proposed Low-complexity Joint Beamforming and Trajectory Optimization Algorithm}\label{section:3}
Problem $P_{0}$ is difficult to be solved optimally due to the non-concave objective function and constraints. Given the UAV's trajectory $\{\mathbf{q}_{n},\forall n\}$, the optimization of the beamforming variables $\{\mathbf{w}_{k,n},\mathbf{v}_{k,n},\mathbf{r}_{j,n},\mathbf{u}_{j,n},\forall k,j,n\}$ over $N$ time slots can be decoupled. Therefore, problem $P_{0}$ with respect to the beamforming variables can be equivalently decomposed into $N$ subproblems and solved by solving $N$ subproblems (one subproblem for one time slot). In addition, even though the beamforming variables are fixed, problem $P_{0}$ with respect to $\{\mathbf{q}_{n},\forall n\}$ is still difficult to be solved optimally since the trajectory variables $\{\mathbf{q}_{n},\forall n\}$ are involved in the channel vectors in an extraordinarily complicated form according to equations (\ref{channel_1}) and (\ref{channel_2}). Therefore, considering the highly decoupling of the beamforming variables over $N$ time slots and highly difficulty of directly optimizing the UAV trajectory, we first propose an efficient alternating optimization algorithm to design the communication and sensing beamforming for a given UAV location, and then propose a low-complexity joint beamforming and UAV trajectory optimization algorithm to find the sub-optimal solution.


\subsection{Transmit and Receive Beamforming Design}
In this subsection, we focus on the optimization of the beamforming variables at time slot $n$ for a given UAV location $\mathbf{q}_{n}$. For notational convenience, we will omit the subscript $n$ in all the notations in the rest of this subsection. After omitting the subscript $n$, the beamforming variables $\{\mathbf{w}_{k,n}, \mathbf{v}_{k,n},\mathbf{r}_{j,n},\mathbf{u}_{j,n},\forall k,j,n\}$ become $\{\mathbf{w}_{k}, \mathbf{v}_{k},\mathbf{r}_{j},\mathbf{u}_{j},\forall k,j\}$. Since the proposed AISAC mechanism enables sensing on demand during communication, we need to optimize the communication and sensing beamforming $\{\mathbf{w}_{k},\mathbf{v}_{k},\mathbf{r}_{j},\mathbf{u}_{j},\forall k,j\}$ if the UAV enables sensing (i.e., $\psi_{n}=1$), otherwise we only need to optimize the communication beamforming $\{\mathbf{w}_{k},\mathbf{v}_{k}, \forall k\}$. The optimization when $\psi_{n}=0$ can be recognized as a special case of the optimization with $\psi_{n}=1$. Therefore, for a given UAV location $\mathbf{q}_{n}$, we optimize $\{\mathbf{w}_{k}, \mathbf{v}_{k},\mathbf{r}_{j},\mathbf{u}_{j},\forall k,j\}$, which corresponds to solving the following problem
\begin{subequations} \label{P_{1}}
\begin{align}
&{P_{1}:}\underset{\{\mathbf{w}_{k}, \mathbf{v}_{k},\mathbf{r}_{j},\mathbf{u}_{j},\forall k,j\}}
{\mathrm{max}}\begin{aligned}&\sum_{k=1}^{K}\big(R_{k}^{\mathrm{DL}}+ R_{k}^{\mathrm{UL}}\big)\end{aligned} \notag
\\& \quad \mathrm{s.t.} \quad \psi\gamma_{j}^{\mathrm{Sens}}\geq \psi\gamma_{\mathrm{th}}^{\mathrm{Sens}},\forall j,\label{P1_a}
\\& \quad\quad\quad \gamma_{k}^{\mathrm{DL}}\geq\gamma_{\mathrm{th}}^{\mathrm{DL}},\forall k,\label{P1_b}
\\& \quad\quad\quad \gamma_{k}^{\mathrm{UL}}\geq\gamma_{\mathrm{th}}^{\mathrm{UL}},\forall k,\label{P1_c}
\\& \quad\quad\quad \sum_{k=1}^{K}\big\|\mathbf{w}_{k}\big\|^{2}
+\psi\sum_{j=1}^{J}\big\|\mathbf{r}_{j}\big\|^{2}\leq P_{\mathrm{max}}^{\mathrm{uav}},\label{P1_d}
\\& \quad\quad\quad \big\|\mathbf{v}_{k}\big\|^{2}=1,
\psi\big\|\mathbf{u}_{j}\big\|^{2}=\psi, \forall k,j.\label{P1_e}
\end{align}
\end{subequations}

Problem $P_{1}$ is difficult to solve due to the non-convexity of the objective function and constraints. To make the problem $P_{1}$ more tractable, we first define $\mathbf{W}_{k}=\mathbf{w}_{k}\mathbf{w}_{k}^{H}$, $\mathbf{V}_{k}=\mathbf{v}_{k}\mathbf{v}_{k}^{H}$,
$\mathbf{R}_{j} = \mathbf{r}_{j}\mathbf{r}_{j}^{H}$, and $\mathbf{U}_{j}=\mathbf{u}_{j}\mathbf{u}_{j}^{H}$, where $\mathbf{W}_{k},\mathbf{V}_{k},\mathbf{R}_{j},\mathbf{U}_{j}\succeq \mathbf{0}$, and $\mathrm{rank}(\mathbf{W}_{k})=\mathrm{rank}(\mathbf{V}_{k})
=\mathrm{rank}(\mathbf{R}_{j})=\mathrm{rank}(\mathbf{U}_{j})=1$. By replacing $\mathbf{w}_{k}\mathbf{w}_{k}^{H}$, $\mathbf{v}_{k}\mathbf{v}_{k}^{H}$, $\mathbf{r}_{j}\mathbf{r}_{j}^{H}$, and $\mathbf{u}_{j}\mathbf{u}_{j}^{H}$ with $\mathbf{W}_{k}$, $\mathbf{V}_{k}$, $\mathbf{R}_{j}$, and $\mathbf{U}_{j}$, respectively, problem $P_{1}$ is recast as the following problem
\begin{subequations} \label{P_1-1}
\begin{align}
&{P_{2}:~} \underset{\mathcal{Z}}{\mathrm{max}}~ \begin{aligned}& \sum_{k=1}^{K}\bigg(\Omega_{k}^{\mathrm{DL}}(\mathcal{Z})
+\Omega_{k}^{\mathrm{UL,1}}(\mathcal{Z})
+\Omega_{k}^{\mathrm{UL,2}}(\mathcal{Z})\bigg)\end{aligned} \notag
\\& \quad \mathrm{s.t.}\quad \psi G_{p}\Phi_{j,j}^{\mathrm{Sens}}(\mathcal{Z})\geq \psi\gamma_{\mathrm{th}}^{\mathrm{Sens}}\big(\sum_{i\neq j}^{J}\Phi_{j,i}^{\mathrm{Sens}}(\mathcal{Z})
+\sigma_{\mathrm{Sens}}^{2}\big),\forall j,\label{P1_1a}
\\& \quad\quad\quad \mathrm{tr}\big(\mathbf{H}_{k}^{\mathrm{DL}}\mathbf{W}_{k}\big)
\geq \gamma_{\mathrm{th}}^{\mathrm{DL}}\Phi_{k}^{\mathrm{DL}}
(\mathcal{Z}),\forall k,\label{P1_1b}
\\& \quad\quad\quad p_{k}\mathrm{tr}\big(\mathbf{H}_{k}^{\mathrm{UL}}\mathbf{V}_{k}\big)
\geq \gamma_{\mathrm{th}}^{\mathrm{UL}}\Phi_{k}^{\mathrm{UL}}
(\mathcal{Z}),\forall k,\label{P1_1c}
\\& \quad\quad\quad \sum_{k=1}^{K}\mathrm{tr}\big(\mathbf{W}_{k}\big)
+\psi\sum_{j=1}^{J}\mathrm{tr}\big(\mathbf{R}_{j}\big)\leq P_{\mathrm{max}}^{\mathrm{uav}},\label{P1_1d}
\\& \quad\quad\quad \mathbf{W}_{k},\mathbf{V}_{k}, \psi\mathbf{R}_{j}, \psi\mathbf{U}_{j}\succeq \mathbf{0}, \forall k,j,\label{P1_1e}
\\& \quad\quad\quad \mathrm{tr}(\mathbf{V}_{k})=1, \psi\mathrm{tr}(\mathbf{U}_{j})=\psi,\forall k,j,\label{P1_1f}
\\& \quad\quad\quad \mathrm{rank}(\mathbf{W}_{k})=\mathrm{rank}(\mathbf{V}_{k})=1, \psi\mathrm{rank}(\mathbf{R}_{j})=\psi\mathrm{rank}(\mathbf{U}_{j})=\psi,\forall k,j,\label{P1_1g}
\end{align}
\end{subequations}
where $\mathcal{Z}=\{\mathbf{W}_{k}, \mathbf{V}_{k}, \mathbf{R}_{j}, \mathbf{U}_{j},\forall k,j\}$, $\mathbf{H}_{k}^{\mathrm{DL}}=\mathbf{h}_{k}^{\mathrm{DL}}(\mathbf{h}_{k}^{\mathrm{DL}})^{H}$, $\mathbf{H}_{k}^{\mathrm{UL}}=\mathbf{h}_{k}^{\mathrm{UL}}(\mathbf{h}_{k}^{\mathrm{UL}})^{H}$, $\Omega_{k}^{\mathrm{DL}}(\mathcal{Z})=
t^{\mathrm{DL}}\log(1+\frac{\mathrm{tr}(\mathbf{H}_{k}^{\mathrm{DL}}\mathbf{W}_{k})}
{\Phi_{k}^{\mathrm{DL}}(\mathcal{Z})})$, $\Omega_{k}^{\mathrm{UL,1}}(\mathcal{Z})=
t^{\mathrm{UL,1}}\log(1+\frac{p_{k}\mathrm{tr}(\mathbf{H}_{k}^{\mathrm{UL}}\mathbf{V}_{k})}
{\Phi_{k}^{\mathrm{UL}}(\mathcal{Z})|_{\xi=1}})$, $\Omega_{k}^{\mathrm{UL,2}}(\mathcal{Z})=
t^{\mathrm{UL,2}}\log(1+\frac{p_{k}\mathrm{tr}(\mathbf{H}_{k}^{\mathrm{UL}}\mathbf{V}_{k})}
{\Phi_{k}^{\mathrm{UL}}(\mathcal{Z})|_{\xi=0}})$, $\Phi_{k}^{\mathrm{DL}}(\mathcal{Z})
=\sum_{i\neq k}^{K}\mathrm{tr}(\mathbf{H}_{k}^{\mathrm{DL}}\mathbf{W}_{i})
+\psi\sum_{j=1}^{J}\mathrm{tr}(\mathbf{H}_{k}^{\mathrm{DL}}\mathbf{R}_{j})
+\sigma_{\mathrm{DL}}^{2}$, $\Phi_{k}^{\mathrm{UL}}(\mathcal{Z})
=\sum_{i\neq k}^{K}p_{i}\mathrm{tr}(\mathbf{H}_{i}^{\mathrm{UL}}\mathbf{V}_{k})
+\xi\sum_{j=1}^{J}\mathrm{tr}(\mathbf{G}_{j}^{H}\mathbf{V}_{k}\mathbf{G}_{j}
(\sum_{i=1}^{J}\mathbf{R}_{j}+\sum_{i=1}^{K}\mathbf{W}_{i}))+\sigma_{\mathrm{UL}}^{2}$,
and $\Phi_{j,i}^{\mathrm{Sens}}(\mathcal{Z})=\mathrm{tr}(\mathbf{U}_{j}\mathbf{G}_{i}(\mathbf{R}_{i}
+\sum_{k=1}^{K}\mathbf{W}_{k})\mathbf{G}_{i}^{H})$. $\Phi_{k}^{\mathrm{UL}}(\mathcal{Z})|_{\xi=1}$ and $\Phi_{k}^{\mathrm{UL}}(\mathcal{Z})|_{\xi=0}$ denote the function $\Phi_{k}^{\mathrm{UL}}(\mathcal{Z})$ conditioned on $\xi=1$ and $\xi=0$, respectively. Note that constraints (\ref{P1_a})-(\ref{P1_d}) have been re-expressed as constraints (\ref{P1_1a})-(\ref{P1_1d}).

In problem $P_{2}$, the objective function is still non-convex and the optimization variables $\mathcal{W}=\{\mathbf{W}_{k},\forall k\}$, $\mathcal{V}=\{\mathbf{V}_{k},\forall k\}$, $\mathcal{R}=\{\mathbf{R}_{j},\forall j\}$, and $\mathcal{U}=\{\mathbf{U}_{j},\forall j\}$ are highly coupled, thus optimizing problem $P_{2}$ is still challenging. To efficiently solve this problem, we first decompose problem $P_{2}$ into multiple subproblems, and then propose an alternating optimization algorithm to solve them, as detailed in the sequel.

\subsubsection{Transmit Beamforming Design for DL Communication}
For given $\mathcal{V}$, $\mathcal{R}$ and $\mathcal{U}$, we focus on optimizing $\mathcal{W}$, which corresponds to solving the following problem
\begin{subequations} \label{SP_1}
\begin{align}
&{P_{3}:~} \underset{\mathcal{W}}{\mathrm{max}}~ \begin{aligned}& \sum_{k=1}^{K}\bigg(\Omega_{k}^{\mathrm{DL}}(\mathcal{Z})
+\Omega_{k}^{\mathrm{UL,1}}(\mathcal{Z})\bigg)
\notag \end{aligned}
\\& \quad \mathrm{s.t.}\quad (\ref{P1_1a})-(\ref{P1_1d}),\label{SP1_a}
\\& \quad\quad\quad \mathbf{W}_{k}\succeq \mathbf{0},\forall k,\label{SP1_b}
\\& \quad\quad\quad \mathrm{rank}(\mathbf{W}_{k})=1,\forall k.\label{SP1_c}
\end{align}
\end{subequations}
It can be seen that problem $P_{3}$ is a sum-of-functions-of-ratio problem. To deal with this problem, we first apply the Lagrangian dual transform method \cite{Fractional_Programming} to equivalently transform problem $P_{3}$ into the following problem
\begin{subequations} \label{SP_{1-1}}
\begin{align}
&{P_{4}:~} \underset{\mathcal{W},\bm{\mathcal{\delta}}}{\mathrm{max}}  \sum_{k=1}^{K}\bigg(t^{\mathrm{DL}}\big(\log(1+\delta_{k}^{\mathrm{DL}})-\delta_{k}^{\mathrm{DL}}\big)
+\frac{C_{k}^{\mathrm{DL}}(\mathcal{W},\bm{\mathcal{\delta}})}
{D_{k}^{\mathrm{DL}}(\mathcal{W},\bm{\mathcal{\delta}})}
\notag\\&~~~~~~~~~~~~~~~~~+t^{\mathrm{UL,1}}\big(\log(1+\delta_{k}^{\mathrm{UL}})-\delta_{k}^{\mathrm{UL}}\big)
+\frac{C_{k}^{\mathrm{UL}}(\mathcal{W},\bm{\mathcal{\delta}})}
{D_{k}^{\mathrm{UL}}(\mathcal{W},\bm{\mathcal{\delta}})}\bigg) \notag
\\& \quad \mathrm{s.t.}\quad (\ref{P1_1a})-(\ref{P1_1d}),(\ref{SP1_b}),(\ref{SP1_c}), \label{SP1_1a}
\end{align}
\end{subequations}
where $C_{k}^{\mathrm{DL}}(\mathcal{W},\bm{\mathcal{\delta}})= t^{\mathrm{DL}}\big(1+\delta_{k}^{\mathrm{DL}}\big)
\mathrm{tr}\big(\mathbf{H}_{k}^{\mathrm{DL}}\mathbf{W}_{k}\big)$,
$C_{k}^{\mathrm{UL}}(\mathcal{W},\bm{\mathcal{\delta}})=
t^{\mathrm{UL,1}}(1+\delta_{k}^{\mathrm{UL}})
p_{k}\mathrm{tr}\big(\mathbf{H}_{k}^{\mathrm{UL}}\mathbf{V}_{k}\big)$,
$D_{k}^{\mathrm{DL}}(\mathcal{W},\bm{\mathcal{\delta}})=
\mathrm{tr}(\mathbf{H}_{k}^{\mathrm{DL}}\mathbf{W}_{k})
+\Phi_{k}^{\mathrm{DL}}(\mathcal{Z})$, $D_{k}^{\mathrm{UL}}(\mathcal{W},\bm{\mathcal{\delta}})=
p_{k}\mathrm{tr}(\mathbf{H}_{k}^{\mathrm{UL}}\mathbf{V}_{k})
+\Phi_{k}^{\mathrm{UL}}(\mathcal{Z})|_{\xi=1}$, and $\bm{\mathcal{\delta}}=\{\delta_{k}^{\mathrm{DL}},\delta_{k}^{\mathrm{UL}}\geq0\}$ is the introduced auxiliary variables. In problem $P_{4}$, all the variables are alternately optimized. When $\mathcal{W}$ is fixed, the optimal $\bm{\mathcal{\delta}}^{\star}$ is obtained by setting the derivative of the objective function with respect to $\bm{\mathcal{\delta}}$ to zero, and given as
\begin{equation}
\begin{aligned}
&\delta_{k}^{\mathrm{DL},\star} = \frac{\mathrm{tr}\big(\mathbf{H}_{k}^{\mathrm{DL}}\mathbf{W}_{k}\big)}
{\Phi_{k}^{\mathrm{DL}}(\mathcal{Z})},
\delta_{k}^{\mathrm{UL},\star} = \frac{p_{k}\mathrm{tr}\big(\mathbf{H}_{k}^{\mathrm{UL}}\mathbf{V}_{k}\big)}
{\Phi_{k}^{\mathrm{UL}}(\mathcal{Z})|_{\xi=1}},\forall k.
\end{aligned}
\end{equation}

Then, we apply the quadratic transform method \cite{Fractional_Programming} to equivalently translate problem $P_{4}$ into a more tractable form
\begin{subequations} \label{SP_{1-2}}
\begin{align}
&{P_{5}:~} \underset{\mathcal{W},\bm{\mathcal{\delta}},\bm{\mathcal{\varepsilon}}}{\mathrm{max}}  \sum_{k=1}^{K}\bigg(t^{\mathrm{DL}}\big(\log(1+\delta_{k}^{\mathrm{DL}})-\delta_{k}^{\mathrm{DL}}\big)
+2\varepsilon_{k}^{\mathrm{DL}}\sqrt{C_{k}^{\mathrm{DL}}(\mathcal{W},\bm{\mathcal{\delta}})}
-(\varepsilon_{k}^{\mathrm{DL}})^{2}D_{k}^{\mathrm{DL}}(\mathcal{W},\bm{\mathcal{\delta}})
\notag\\&~~~~~~~~~~~~~~~~~+t^{\mathrm{UL,1}}\big(\log(1+\delta_{k}^{\mathrm{UL}})-\delta_{k}^{\mathrm{UL}}\big)
+2\varepsilon_{k}^{\mathrm{UL}}\sqrt{C_{k}^{\mathrm{UL}}(\mathcal{W},\bm{\mathcal{\delta}})}
-(\varepsilon_{k}^{\mathrm{UL}})^{2}D_{k}^{\mathrm{UL}}(\mathcal{W},\bm{\mathcal{\delta}})\bigg) \notag
\\& \quad \mathrm{s.t.}\quad (\ref{P1_1a})-(\ref{P1_1d}),(\ref{SP1_b}),(\ref{SP1_c}), \label{SP1_1a}
\end{align}
\end{subequations}
where $\bm{\mathcal{\varepsilon}}=\{
\varepsilon_{k}^{\mathrm{DL}},\varepsilon_{k}^{\mathrm{UL}}\geq0\}$ is the introduced auxiliary variables. In problem $P_{5}$, $\mathcal{W}$, $\bm{\mathcal{\delta}}$, and $\bm{\mathcal{\varepsilon}}$ are alternately optimized by using the proposed alternating optimization method summarized in Algorithm \ref{alg:Iterative_1}. When $\mathcal{W}$ and $\bm{\mathcal{\delta}}$ are held fixed, the optimal $\bm{\mathcal{\varepsilon}}^{\star}$ is obtained by setting the derivative of the objective function with respect to $\bm{\mathcal{\varepsilon}}$ to zero, and given as
\begin{equation}
\begin{aligned}
&\varepsilon_{k}^{\mathrm{DL},\star} = \frac{\sqrt{C_{k}^{\mathrm{DL}}(\mathcal{W},\bm{\mathcal{\delta}})}}
{D_{k}^{\mathrm{DL}}(\mathcal{W},\bm{\mathcal{\delta}})},
\varepsilon_{k}^{\mathrm{UL},\star} = \frac{\sqrt{C_{k}^{\mathrm{UL}}(\mathcal{W},\bm{\mathcal{\delta}})}}
{D_{k,n}^{\mathrm{UL}}(\mathcal{W},\bm{\mathcal{\delta}})},\forall k.
\end{aligned}
\end{equation}

In problem $P_{5}$, for fixed $\bm{\mathcal{\delta}}$ and $\bm{\mathcal{\varepsilon}}$, the objective function with respect to $\mathcal{W}$ is concave. However, problem $P_{5}$ is still non-convex due to the non-convex rank constraints in (\ref{SP1_c}). We relax the rank constraints, so that problem $P_{5}$ over $\mathcal{W}$ becomes a standard convex optimization problem, which can be solved optimally by the convex optimization solver CVX. For the rank constraints, when there is no guarantee that the obtained optimal $\mathcal{W}^{\star}$ is of rank-one, the Gaussian randomization technique \cite{Semidefinite_Relaxation} can be utilized to obtain an approximated solution that satisfies the rank-one constraints. Fortunately, by checking the Karush-Kuhn-Tucker (KKT) optimality conditions of the relaxed version of problem $P_{5}$ and applying the similar analysis as in the Appendix of \cite{9481926Rank,9570143Rank}, there always exists an optimal rank-one solution $\mathcal{W}^{\star}$ to problem $P_{5}$, and thus the Gaussian randomization is not needed for solving problem $P_{5}$, in which the detailed proof is omitted due to space limitations.

\begin{algorithm} [!t]
\setlength\abovedisplayskip{1pt}
\setlength\belowdisplayskip{1pt}
	\caption{Transmit Beamforming Design for DL Communication.}\label{alg:Iterative_1}
	\KwIn{$\mathcal{W}^{0}$, iteration number $l_{1}=1$, and maximum iteration number $l_{1}^{\mathrm{max}}$.}

       \Repeat{The objective function value converges or $l_{1}>l_{1}^{\mathrm{max}}$}
       {
       Given $\mathcal{W}^{(l_{1}-1)}$, update the auxiliary variables $\bm{\delta}^{(l_{1})}=\{\delta_{k}^{\mathrm{DL},(l_{1})},\delta_{k}^{\mathrm{UL},(l_{1})},\forall k\}$;\\

       Given $\mathcal{W}^{(l_{1}-1)}$ and $\bm{\delta}^{(l_{1})}$, update the auxiliary variables $\bm{\varepsilon}^{(l_{1})}=\{\varepsilon_{k}^{\mathrm{DL},(l_{1})},\varepsilon_{k}^{\mathrm{UL},(l_{1})},\forall k\}$;\\

       Given $\bm{\delta}^{(l_{1})}$ and $\bm{\varepsilon}^{(l_{1})}$, update the optimization variable $\mathcal{W}^{(l_{1})}$;\\

       $l_{1}=l_{1}+1$;

        }

		\KwOut{$\mathcal{W}^{\star}$.}
\end{algorithm}

\subsubsection{Receive Beamforming Design for UL Communication}
For given $\mathcal{W}$, $\mathcal{R}$, and $\mathcal{U}$, problem $P_{2}$ over $\mathcal{V}$ can be formulated as
\begin{subequations} \label{SP_{2}}
\begin{align}
&{P_{6}:~} \underset{\mathcal{V}}{\mathrm{max}} \begin{aligned}&\sum_{k=1}^{K}\bigg(\Omega_{k}^{\mathrm{DL}}(\mathcal{Z})
+\Omega_{k}^{\mathrm{UL,1}}(\mathcal{Z})\bigg)\end{aligned} \notag
\\& \quad \mathrm{s.t.}\quad (\ref{P1_1c}),\label{SP2_a}
\\& \quad\quad\quad \mathbf{V}_{k}\succeq \mathbf{0},\mathrm{tr}(\mathbf{V}_{k})=1,\forall k,\label{SP2_b}
\\& \quad\quad\quad \mathrm{rank}(\mathbf{V}_{k})=1,\forall k.\label{SP2_c}
\end{align}
\end{subequations}
Problem $P_{6}$ is also a sum-of-functions-of-ratio problem, which has a similar form to problem $P_{3}$. As a result, after relaxing the rank constraints, problem $P_{6}$ can be solved similarly as solving problem $P_{3}$, for which the details are omitted for brevity. Besides, by checking the KKT optimality conditions of the relaxed version of problem $P_{6}$, $\mathrm{rank}(\mathbf{V}_{k})=1$ always holds.

\subsubsection{Transmit Beamforming Design for Sensing}
In problem $P_{2}$, the transmit beamforming for sensing is required if and only if the UAV decides to enable sensing. For given $\mathcal{W}$, $\mathcal{V}$, and $\mathcal{U}$, problem $P_{2}$ over $\mathcal{R}$ can be formulated as
\begin{subequations} \label{SP_3}
\begin{align}
&{P_{7}:~} \underset{\mathcal{R}}{\mathrm{max}} \begin{aligned}& \sum_{k=1}^{K}\bigg(\Omega_{k}^{\mathrm{DL}}(\mathcal{Z})
+\Omega_{k}^{\mathrm{UL,1}}(\mathcal{Z})\bigg)
\end{aligned} \notag
\\& \quad \mathrm{s.t.}\quad (\ref{P1_1a})-(\ref{P1_1d}),\label{SP3_a}
\\& \quad\quad\quad \mathbf{R}_{j}\succeq \mathbf{0},\forall j, \label{SP3_b}
\\& \quad\quad\quad \mathrm{rank}(\mathbf{R}_{j})=1,\forall j. \label{SP3_c}
\end{align}
\end{subequations}
Although $\mathcal{W}$, $\mathcal{V}$, and $\mathcal{U}$ are fixed, problem $P_{7}$ is still non-convex due to the non-convex objective function and rank-one constraints. To tackle this non-convex problem, we introduce auxiliary variables $\bm{\mathcal{\chi}}=\{\chi_{k}^{\mathrm{DL}}=\mathrm{tr}\big(\mathbf{H}_{k}^{\mathrm{DL}}\mathbf{W}_{k}\big)/
\Phi_{k}^{\mathrm{DL}}(\mathcal{Z}),\chi_{k}^{\mathrm{UL}}=p_{k}\mathrm{tr}\big(\mathbf{H}_{k}^{\mathrm{UL}}\mathbf{V}_{k}\big)/
\Phi_{k}^{\mathrm{UL}}(\mathcal{Z})|_{\xi=1}\}$, and then transform problem $P_{7}$ into the following problem
\begin{subequations} \label{SP_3-1}
\begin{align}
&{P_{8}:~} \underset{\mathcal{R},\bm{\mathcal{\chi}}}{\mathrm{max}} \begin{aligned}& \sum_{k=1}^{K}\big(t^{\mathrm{DL}}\log(1+\chi_{k}^{\mathrm{DL}})
+t^{\mathrm{UL,1}}\log(1+\chi_{k}^{\mathrm{UL}})\big)\end{aligned} \notag
\\& \quad \mathrm{s.t.}\quad (\ref{P1_1a})-(\ref{P1_1d}),(\ref{SP3_b}),(\ref{SP3_c}),\label{SP3_1a}
\\& \quad\quad\quad \frac{\mathrm{tr}\big(\mathbf{H}_{k}^{\mathrm{DL}}\mathbf{W}_{k}\big)}
{\Phi_{k}^{\mathrm{DL}}(\mathcal{Z})} \geq \chi_{k}^{\mathrm{DL}},\forall k,\label{SP3_1b}
\\& \quad\quad\quad \frac{p_{k}\mathrm{tr}\big(\mathbf{H}_{k}^{\mathrm{UL}}\mathbf{V}_{k}\big)}
{\Phi_{k}^{\mathrm{UL}}(\mathcal{Z})|_{\xi=1}} \geq \chi_{k}^{\mathrm{UL}},\forall k.\label{SP3_1c}
\end{align}
\end{subequations}

Before solving problem $P_{8}$, we rewrite the constraints (\ref{SP3_1b}) and (\ref{SP3_1c}) as
\begin{align}
\mathrm{tr}\big(\mathbf{H}_{k}^{\mathrm{DL}}\mathbf{W}_{k}\big)
\geq \chi_{k}^{\mathrm{DL}}\Phi_{k}^{\mathrm{DL}}
(\mathcal{Z}),\forall k,\label{SP3_1d}\\
p_{k}\mathrm{tr}\big(\mathbf{H}_{k}^{\mathrm{UL}}\mathbf{V}_{k}\big)
\geq \chi_{k}^{\mathrm{UL}}\Phi_{k}^{\mathrm{UL}}
(\mathcal{Z})|_{\xi=1},\forall k. \label{SP3_1e}
\end{align}

Since the right hand terms in inequations (\ref{SP3_1d}) and (\ref{SP3_1e}) are quasi-concave, the constraints (\ref{SP3_1d}) and (\ref{SP3_1e}) are still non-convex. To deal with these non-convex constraints, we invoke the convex upper bound approximation method \cite{upper_bound_approximation} and transform the constraints (\ref{SP3_1d}) and (\ref{SP3_1e}) into the following inequations
\begin{align}
\mathrm{tr}\big(\mathbf{H}_{k}^{\mathrm{DL}}\mathbf{W}_{k}\big)
\geq \frac{\big(\Phi_{k}^{\mathrm{DL}}
(\mathcal{Z})\big)^{2}}{2\Theta_{k}^{\mathrm{DL}}}
+\frac{\Theta_{k}^{\mathrm{DL}}(\chi_{k}^{\mathrm{DL}})^{2}}{2},\forall k,\label{SP3_1f}\\
p_{k}\mathrm{tr}\big(\mathbf{H}_{k}^{\mathrm{UL}}\mathbf{V}_{k}\big)
\geq \frac{\big(\Phi_{k}^{\mathrm{UL}}(\mathcal{Z})|_{\xi=1}\big)^{2}}{2\Theta_{k}^{\mathrm{UL}}}
+\frac{\Phi_{k}^{\mathrm{UL}}(\chi_{k}^{\mathrm{UL}})^{2}}{2},\forall k, \label{SP3_1g}
\end{align}
where the right hand terms in inequations (\ref{SP3_1f}) and (\ref{SP3_1g}) are convex. Equalities in constraints (\ref{SP3_1f}) and (\ref{SP3_1g}) will always hold if $\Theta_{k}^{\Lambda}=\Phi_{k}^{\Lambda}(\mathcal{Z})/\chi_{k,n}^{\Lambda}$, where $\Lambda\in\{\mathrm{DL,UL}\}$. In the $p$-th iteration, $\Theta_{k}^{\Lambda}$ is updated by $(\Theta_{k}^{\Lambda})^{(p)}=\Phi_{k}^{\Lambda}(\mathcal{Z}^{(p-1)})
/(\chi_{k}^{\Lambda})^{(p-1)}$.

Finally, problem $P_{8}$ is transformed to the following problem
\begin{subequations} \label{SP_3-2}
\begin{align}
&{P_{9}:~} \underset{\mathcal{R},\bm{\mathcal{\chi}}}{\mathrm{max}} \begin{aligned}& \sum_{k=1}^{K}\big(t^{\mathrm{DL}}\log(1+\chi_{k}^{\mathrm{DL}})
+t^{\mathrm{UL,1}}\log(1+\chi_{k}^{\mathrm{UL}})\big)\end{aligned} \notag
\\& \quad \mathrm{s.t.}\quad (\ref{P1_1a})-(\ref{P1_1d}),(\ref{SP3_b}),(\ref{SP3_c}),(\ref{SP3_1f}),(\ref{SP3_1g}). \label{SP3_2a}
\end{align}
\end{subequations}

It is noted that after relaxing the non-convex rank-one constraints (\ref{SP3_c}), problem $P_{9}$ is convex with respect to $\mathcal{R}$ and $\bm{\mathcal{\chi}}$, which can be solved efficiently with the standard convex optimization algorithms or software such as CVX. By checking the KKT optimality conditions of the relaxed version of problem $P_{9}$, the optimal $\mathcal{R}^{\star}$ is of rank-one.

Based on the above processing, the overall algorithm for solving the original non-convex problem $P_{7}$ is summarized in Algorithm \ref{alg:Iterative_2}. Since the objective value produced by Algorithm \ref{alg:Iterative_2} is non-decreasing over iterations and upper bounded by a finite value, Algorithm \ref{alg:Iterative_2} is guaranteed to converge.
\begin{algorithm} [!t]
\setlength\abovedisplayskip{1pt}
\setlength\belowdisplayskip{1pt}
	\caption{Transmit Beamforming Design for Sensing.}\label{alg:Iterative_2}
	\KwIn{$\mathcal{R}^{0}$, $\bm{\mathcal{\chi}}^{0}$, iteration number $l_{2}=1$ and maximum iteration number $l_{2}^{\mathrm{max}}$.}

       \Repeat{The objective function value converges or $l_{2}>l_{2}^{\mathrm{max}}$}
       {
       Given $\mathcal{R}^{(l_{2}-1)}$ and $\bm{\mathcal{\chi}}^{(l_{2}-1)}$, calculate $\bm{\Theta}^{(l_{2})}=\{(\Theta_{k}^{DL})^{(l_{2})},(\Theta_{k}^{UL})^{(l_{2})},\forall k\}$;

       Given $\bm{\Theta}^{(l_{2})}$, update $\mathcal{R}^{(l_{2})}$ and $\bm{\mathcal{\chi}}^{(l_{2})}$; \\

       $l_{2}=l_{2}+1$;\\
        }

		\KwOut{$\mathcal{R}^{\star}$.}
\end{algorithm}

\subsubsection{Receive Beamforming Design for Sensing}
In problem $P_{2}$, the receive beamforming for sensing is required if and only if the UAV enables sensing. For given $\mathcal{W}$, $\mathcal{V}$, and $\mathcal{R}$, problem $P_{2}$ is recast as follows
\begin{subequations} \label{SP_{4}}
\begin{align}
&{P_{10}:~}
\mathrm{Find}~
\begin{aligned}\mathcal{U}&\end{aligned} \notag
\\& \quad \mathrm{s.t.}\quad (\ref{P1_1a}),\label{SP4_a}
\\& \quad\quad\quad \mathbf{U}_{j}\succeq \mathbf{0}, \mathrm{tr}\big(\mathbf{U}_{j}\big) = 1,\forall j,\label{SP4_b}
\\& \quad\quad\quad \mathrm{rank}(\mathbf{U}_{j})=1,\forall j.\label{SP4_c}
\end{align}
\end{subequations}
Problem $P_{10}$ focuses on finding the feasible $\mathcal{U}$ and is still a non-convex problem, due to the non-convex rank-one constraints in (\ref{SP4_c}). To deal with the non-convexity, we again relax the rank-one constraints, and then problem $P_{10}$ becomes a standard semi-definite programming problem, thus can be solved optimally by solver CVX. By checking the KKT optimality conditions of the relaxed version of problem $P_{10}$, the optimal solution $\mathcal{U}^{\star}$ to problem $P_{10}$ is always of rank-one.

The alternating optimization algorithm for AISAC beamforming designs is detailed in Algorithm \ref{alg:Iterative_3}, which optimizes the different beamforming variables alternately until the objective value converges or the maximum iteration number is reached. Due to the monotonic convergence of Algorithm \ref{alg:Iterative_1} and Algorithm \ref{alg:Iterative_2}, the objective value of problem $P_{2}$ is non-decreasing over iterations, and thus Algorithm \ref{alg:Iterative_3} is guaranteed to converge.
\begin{algorithm} [!t]
\setlength\abovedisplayskip{1pt}
\setlength\belowdisplayskip{1pt}
	\caption{Beamforming Optimization Algorithm for the UAV-enabled AISAC System.}\label{alg:Iterative_3}
	\KwIn{$\mathcal{W}^{(0)}$, $\mathcal{V}^{(0)}$, $\mathcal{R}^{(0)}$, $\mathcal{U}^{(0)}$, $\mathbf{q}_{n}$, iteration number $l_{3}=1$, and maximum iteration number $l_{3}^{\mathrm{max}}$.}
    \uIf{The UAV enables ISAC.}{
            \Repeat{The objective function value in problem $P_{1}$ converges or $l_{3}>l_{3}^{\mathrm{max}}$}
           {
            Given $\mathcal{V}^{(l_{3}-1)},\mathcal{R}^{(l_{3}-1)},\mathcal{U}^{(l_{3}-1)}$, update $\mathcal{W}^{(l_{3})}$ via Algorithm \ref{alg:Iterative_1};\\
            Given $\mathcal{W}^{(l_{3})},\mathcal{R}^{(l_{3}-1)},\mathcal{U}^{(l_{3}-1)}$, update $\mathcal{V}^{(l_{3})}$ via the quadratic transform method \cite{Fractional_Programming};\\
            Given $\mathcal{W}^{(l_{3})},\mathcal{V}^{(l_{3})},\mathcal{U}^{(l_{3}-1)}$, update $\mathcal{R}^{(l_{3})}$ via Algorithm \ref{alg:Iterative_2};\\
            Given $\mathcal{W}^{(l_{3})},\mathcal{V}^{(l_{3})},\mathcal{R}^{(l_{3})}$, update $\mathcal{U}^{(l_{3})}$ via solver CVX;\\
            $l_{3}=l_{3}+1$;
            }
}
    \ElseIf{The UAV enables communication.}{
            Given $\mathcal{V}^{(l_{3}-1)}$, update $\mathcal{W}^{(l_{3})}$ via Algorithm \ref{alg:Iterative_1};\\
 Given $\mathcal{W}^{(l_{3})}$, update $\mathcal{V}^{(l_{3})}$ via the quadratic transform method \cite{Fractional_Programming};
           }
		\KwOut{$\mathcal{W}^{\star}$, $\mathcal{V}^{\star}$, $\mathcal{R}^{\star}$, and $\mathcal{U}^{\star}$.}
\end{algorithm}

\subsection{Joint Beamforming and UAV Trajectory Optimization}
\begin{figure}[!t]
\renewcommand{\figurename}{Fig.}
  \centering
  \includegraphics[width=2in]{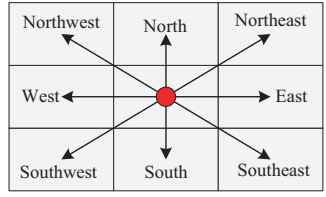}
  \caption{Virtual equally sized rectangular grids. The UAV moves along the  geometric center of the grid, and the distance between two adjacent geometric centers cannot exceed the maximum flying distance within one time slot. Each grid has multiple searchable/reachable directions.}
\label{Fig:2}
\end{figure}

We partition the ground area covered by the UAV into the virtual equally sized rectangular grids\footnote{Note that we take the rectangular grid as an example for simplicity and our proposed methods can also easily extend the shape of the grid to other cases, such as hexagonal.}. Further, we construct a grid map, in which each grid uses the horizontal coordinates of its geometric center as a unique location information identification. Each grid has multiple searchable/reachable directions, as shown in Fig. \ref{Fig:2}, and the distance between the geometric centers of any two adjacent grids cannot exceed the UAV's maximum flying distance within one time slot. The UAV moves along the geometric center of the grid, and either hovers at the grid or moves to one of its reachable grids in each time slot. We aim to need to search the optimal grids/waypoints for maximizing the average system throughput. Although an exhaustive searching method can be used to find the global optimal solution of the trajectory optimization problem, it has intolerable computational complexity. In this subsection, we propose a low-complexity joint beamforming and trajectory optimization algorithm that successively searches for the optimal waypoint from the reachable waypoints until the final location is reached. The details of the proposed algorithm are presented in Algorithm \ref{alg:Iterative_4}.

\begin{algorithm} [!t]
\setlength\abovedisplayskip{1pt}
\setlength\belowdisplayskip{1pt}
	\caption{Low-complexity Joint Beamforming and Trajectory Optimization Algorithm for $\mathbf{P_{0}}$.}\label{alg:Iterative_4}
	\KwIn{Initial location $\mathbf{q}_{I}$, final location $\mathbf{q}_{F}$, time slot number $N$, $n=1$, and table $closelist$.}

\Repeat{$\mathbf{q}_{N}=\mathbf{q}_{F}$}{
Find all reachable waypoints of the waypoint $\mathbf{q}_{n}$;\\

\For{Each reachable waypoint}{
               \uIf{The reachable waypoint is not in $closelist$}{
Run Algorithm \ref{alg:Iterative_5} to obtain the optimal path, calculate $\hat{R}_{n}^{\mathrm{path}}$, and add the reachable waypoint and $\hat{R}_{n}^{\mathrm{path}}$ to $closelist$;\\} \Else{continue;}
}
Select the waypoint with the maximum $\hat{R}_{n}^{\mathrm{path},\star}$ as the optimal waypoint at time slot $(n+1)$, i.e., $\mathbf{q}_{n+1}$;\\
$n=n+1$;
}
		\KwOut{The optimized trajectory.}
\end{algorithm}

\begin{algorithm} [!t]
\setlength\abovedisplayskip{1pt}
\setlength\belowdisplayskip{1pt}
	\caption{Optimal Path Searching Algorithm.}\label{alg:Iterative_5}
	\KwIn{Table $closelist$, time slot number $N$, time slot $n$, waypoint $\mathbf{q}_{n}$, and final location $\mathbf{q}_{F}$.}

\Repeat{$\mathbf{q}_{N}=\mathbf{q}_{F}$}{
Find all reachable waypoints of the waypoint $\mathbf{q}_{n}$;\\

\For{Each reachable waypoint}{
                 \uIf{The reachable waypoint is not in $closelist$}{Run Algorithm \ref{alg:Iterative_3} to optimize the beamforming designs, calculate $\hat{R}_{n}^{\mathrm{waypoint}}$ based on the optimized beamforming, and add the reachable waypoint and $\hat{R}_{n}^{\mathrm{waypoint}}$ to $closelist$;}
                   \Else{continue;}}

Select the waypoint with the maximum $\hat{R}_{n}^{\mathrm{waypoint},\star}$ as the optimal waypoint at time slot $n$, i.e., $\mathbf{q}_{n+1}$;\\
$n=n+1;$
}
		\KwOut{The optimal path from the waypoint $\mathbf{q}_{n}$ to the final location $\mathbf{q}_{F}$.}
\end{algorithm}

In Algorithm \ref{alg:Iterative_4}, for the waypoint $\mathbf{q}_{n}$ at time slot $n$, we first find its all reachable waypoints that can reach the final location $\mathbf{q}_{F}$ within $(N-n)$ time slots. Then, we find the optimal paths of all reachable waypoints by running Algorithm \ref{alg:Iterative_5} and compute the average throughput of each optimal path by the equation $\hat{R}_{n}^{\mathrm{path}}=\frac{1}{N-n+1}\sum_{i=n}^{N}\sum_{k=1}^{K}\big(R_{k,i}^{DL}
+R_{k,i}^{UL}\big)$. Finally, we select the waypoint with the maximum average throughput, i.e., $\hat{R}_{n}^{\mathrm{path},\star}$, as the optimal waypoint at time slot $n$. By repeating the above procedure until the optimal waypoint at time slot $N$ is the final location $\mathbf{q}_{F}$, the optimal trajectory can be obtained. Specifically, Algorithm \ref{alg:Iterative_5} finds the optimal path by successively searching the optimal waypoint until the final location $\mathbf{q}_{N}$ is reached, where the optimal waypoint is the waypoint with the maximum average throughput, i.e., $\hat{R}_{n}^{\mathrm{waypoint},\star}$. Different from Algorithm \ref{alg:Iterative_4}, Algorithm \ref{alg:Iterative_5} computes the average throughput by the equation $\hat{R}_{n}^{\mathrm{waypoint}}=\sum_{k=1}^{K}\big(R_{k,n}^{DL}+R_{k,n}^{UL}\big)$. Note that the calculation of $\hat{R}_{n}^{\mathrm{waypoint}}$ happens after running the Algorithm \ref{alg:Iterative_3}. Considering that different waypoints may have the same reachable waypoints, we introduce a table named $closelist$ to record the waypoints that have been calculated in Algorithm \ref{alg:Iterative_4} and Algorithm \ref{alg:Iterative_5}, thus avoiding the redundant computational work and reducing the searching time. During the searching process in both Algorithm \ref{alg:Iterative_4} and Algorithm \ref{alg:Iterative_5}, each waypoint should be checked first if it is in $closelist$. Specifically, in Algorithm \ref{alg:Iterative_4}, if a waypoint is not in $closelist$, we first need to run Algorithm \ref{alg:Iterative_5} for this waypoint to calculate $\hat{R}_{n}^{\mathrm{path}}$, and then add this waypoint and $\hat{R}_{n}^{\mathrm{path}}$ into $closelist$. In Algorithm \ref{alg:Iterative_5}, if a waypoint is not in $closelist$, we first need to run Algorithm \ref{alg:Iterative_3} for optimizing the beamforming and calculate $\hat{R}_{n}^{\mathrm{waypoint}}$, and then add this waypoint and $\hat{R}_{n}^{\mathrm{waypoint}}$ into $closelist$.

In Algorithm \ref{alg:Iterative_3}, we solve all subproblems with the interior point method \cite{Semidefinite_Relaxation,boyd2004convex}. When the UAV enables ISAC at time slot $n$, the complexity of four subproblems can be given as $\mathcal{O}(\log(1/\epsilon)(KQ^{2})^{3.5})$, $\mathcal{O}(\log(1/\epsilon)(KQ^{2})^{3.5})$, $\mathcal{O}(\log(1/\epsilon)(JQ^{2})^{3.5})$, and $\mathcal{O}(\log(1/\epsilon)(JQ^{2})^{3.5})$, respectively, in which only the highest order term is reserved and $\epsilon$ represents the stopping tolerance. Denote $N_{\mathrm{iter}}$ as the iteration number, and then the total complexity of Algorithm \ref{alg:Iterative_3} is $\mathcal{O}(N_{\mathrm{iter}}\log(1/\epsilon)(2(KQ^{2})^{3.5}+2(JQ^{2})^{3.5}))$. Denote $N_{\mathrm{wp}}$ as the maximum number of reachable waypoints for a waypoint, and then the complexity of Algorithm \ref{alg:Iterative_5} can be given as $\mathcal{O}(NN_{\mathrm{wp}}N_{\mathrm{iter}}\log(1/\epsilon)(2(KQ^{2})^{3.5}+2(JQ^{2})^{3.5}))$. Therefore, the complexity of Algorithm \ref{alg:Iterative_4} in the worst case can be given as $\mathcal{O}(N^{2}N_{\mathrm{wp}}N_{\mathrm{iter}}\log(1/\epsilon)((KQ^{2})^{3.5}+(JQ^{2})^{3.5}))$.

\section{Simulation Results}\label{section:4}
We consider a UAV-enabled AISAC system, where $6$ single-antenna UEs are randomly distributed within the $1$ km $\times$ $1$ km ground area partitioned into $20\times20$ equally sized rectangular grids. The $12$-antenna UAV provides communication services for the ground area and senses on demand within the period $T=20$ s. Unless otherwise stated, we consider two extreme cases in this simulation: No sensing and only communication in each time slot (i.e., $\bm{\psi}=\{\psi_{n}, \forall n\}=\mathbf{0}$) and ISAC in each time slot (i.e., $\bm{\psi}=\{\psi_{n}, \forall n\}=\mathbf{1}$). The initial and final horizontal locations of the UAV are set to $(25, 525)$ and $(975, 525)$, respectively. There are three sensing locations, whose horizontal coordinates are $(375, 175)$, $(375, 725)$, and $(775, 275)$, respectively. Unless otherwise stated, the remaining system parameters are set as follows: $N=20$, $\tau=1$ s, $\beta_{0}=-40$ dB, $p_{k}=25$ dBm, $P_{\mathrm{max}}^{\mathrm{UAV}}=37$ dBm, $\sigma_{j}=1$, $\sigma_{\mathrm{DL}}^{2}=\sigma_{\mathrm{UL}}^{2}=\sigma_{\mathrm{Sens}}^{2}=-110$ dBm, $\gamma_{k,n}^{\mathrm{DL}}=\gamma_{k,n}^{\mathrm{UL}}=0.3$ dB, and $\gamma_{j,n}^{\mathrm{Sens}}=3$ dB.
\begin{figure}[!t]
\renewcommand{\figurename}{Fig.}
  \centering
  \includegraphics[width=9cm]{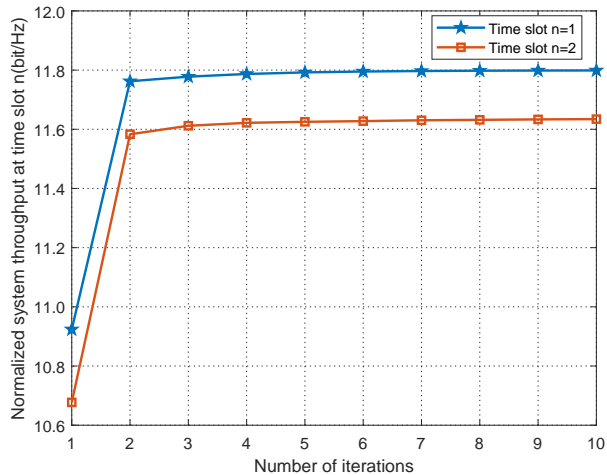}
  \caption{Convergence of the proposed alternating optimization algorithm for beamforming designs.}
\label{Fig:4}
\end{figure}

\subsection{Convergence Properties of the Proposed Beamforming Optimization Algorithm}
To ensure the convergence of the proposed beamforming optimization algorithm for the UAV-enabled AISAC system, we need to show its convergence behavior. Since the beamforming optimization when the UAV only performs communication (i.e., $\bm{\psi}=\mathbf{0}$) can be recognized as a special case of the beamforming optimization when the UAV enables ISAC (i.e., $\bm{\psi}=\mathbf{1}$), we only provide the convergence properties of the proposed algorithm when $\bm{\psi}=\mathbf{1}$. Fig. \ref{Fig:4} plots the average system throughput of two different time slots. One can see that the average system throughput increases drastically at first and then gradually converges to a stable value as the number of iterations increases, thus validating the effectiveness of the proposed optimization algorithm.
\subsection{Trajectory Comparisons of Different Schemes}
\begin{figure}[!t]
\renewcommand{\figurename}{Fig.}
  \centering
  \includegraphics[width=9cm]{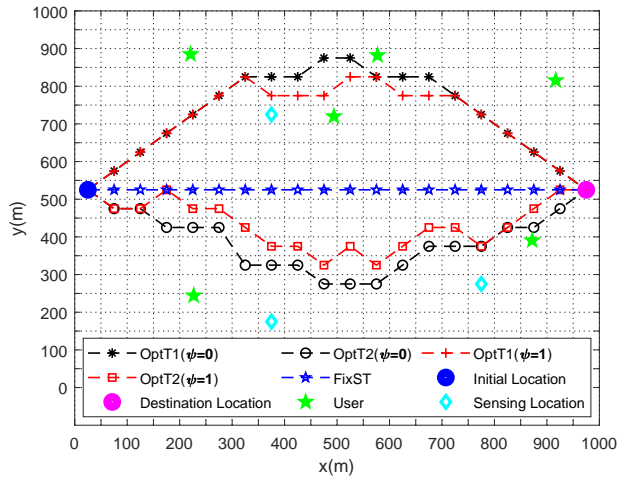}
  \caption{UAV trajectories of different schemes.}
\label{Fig:5}
\end{figure}
In what follows, we compare the UAV trajectories generated by three different schemes: Optimized trajectory generated by Algorithm \ref{alg:Iterative_4} (OptT1); Optimized trajectory generated by Algorithm \ref{alg:Iterative_5} (OptT2); and Fix straight trajectory (FixST). In the FixST scheme, the UAV first flies straightly at the shortest time from the initial location to the optimal location with the maximum average system throughput, then hovers at this optimal location as long as possible, and finally flies straightly toward the final location with the shortest time. All three schemes use Algorithm \ref{alg:Iterative_3} to optimize the beamforming.

In Fig. \ref{Fig:5}, the UAV moves along the geometric center of the grid. It can be seen that, in the OptT1 scheme, the UAV always flies directly from the initial location $\mathbf{q}_{I}$ at the shortest time at the beginning, then slows down to hover above a fixed area as long as possible, and finally flies to the final location $\mathbf{q}_{F}$ with the shortest time. Compared with the other two schemes, in the OptT1 scheme, the UAV always selects a path away from the area with fewer users while maintaining proximity to more users for maximizing the average system throughput. Furthermore, we can observe that when the UAV only performs communication (i.e., $\bm{\psi}=\mathbf{0}$), it has more freedom to adjust its trajectory for maximizing the average system throughput, thus the flying region of the UAV in both OptT1 and OptT2 schemes is relatively large. While when the UAV performs ISAC (i.e., $\bm{\psi}=\mathbf{1}$), the QoS requirements of sensing limit the flying region of the UAV.

\subsection{Performance Comparisons of Different Period and Grid Number}
\begin{figure}[!t]
\renewcommand{\figurename}{Fig.}
\centering
\subfigure[]{
\begin{minipage}{7.8cm}
\includegraphics[width=1\textwidth]{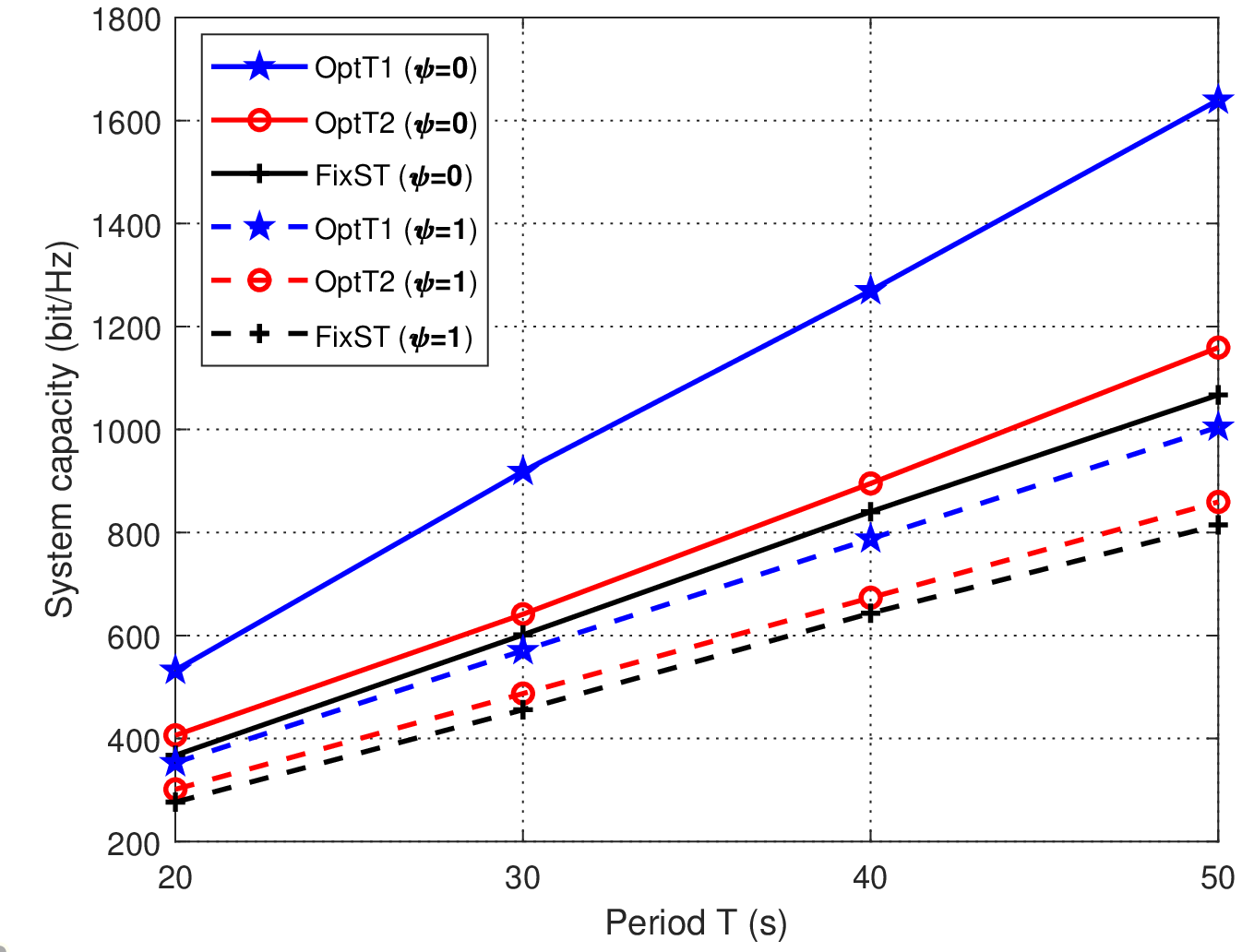}
\label{Fig:6a}
\end{minipage}
}
\subfigure[]{
\begin{minipage}{7.8cm}
\includegraphics[width=1\textwidth]{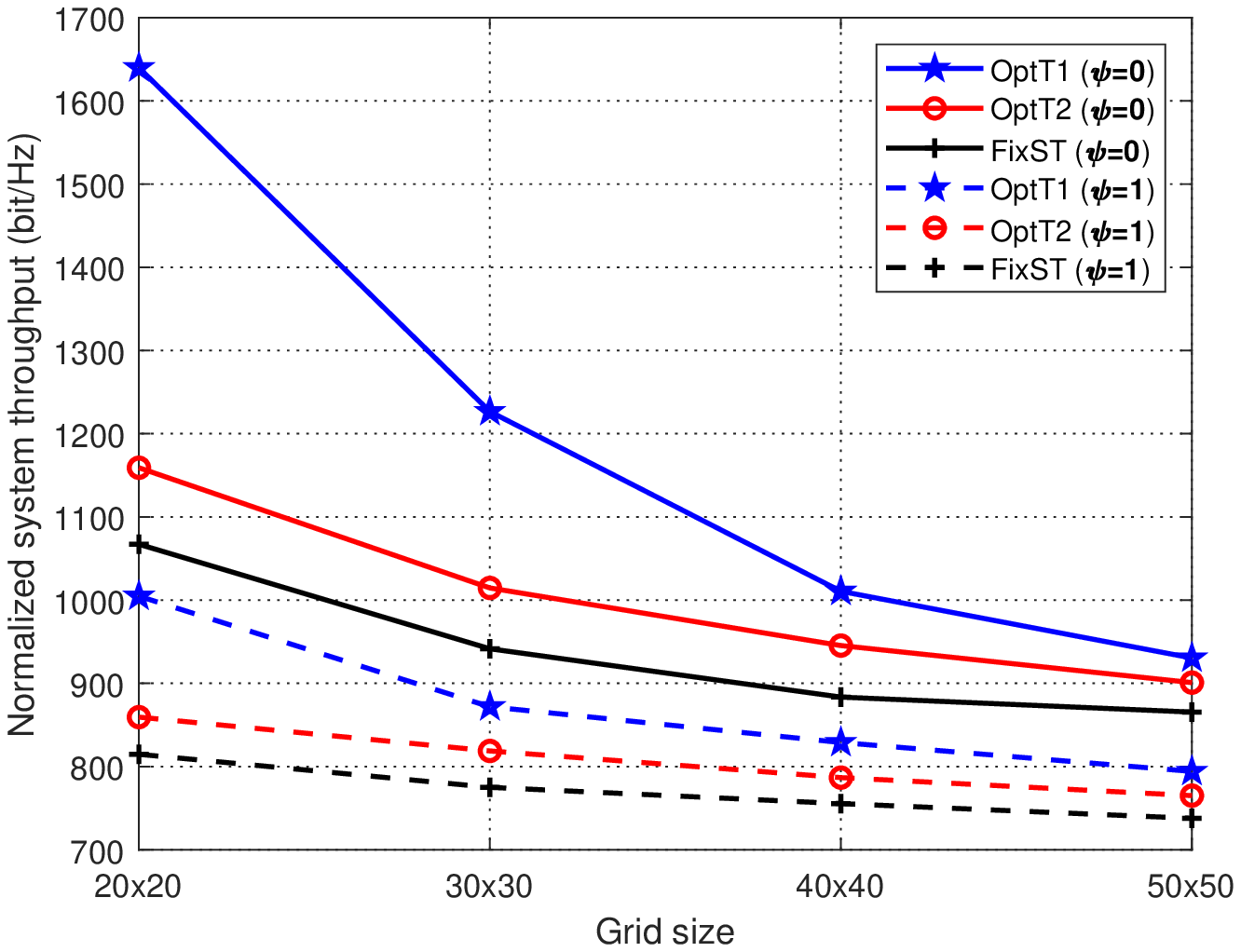}
\label{Fig:6b}
\end{minipage}
}
\caption{(a) Average system throughput of different schemes under different periods. (b) Average system throughput of different schemes under different grid numbers.}
\label{Fig:6}
\end{figure}

To investigate the impact of the period and grid number on the average system throughput, Fig. \ref{Fig:6} shows the average system throughput under different period and grid number settings. In Fig. \ref{Fig:6a}, it can be observed that the average system throughput increases sharply as the value of the period $T$ increases. This is due to the fact that a large period $T$ offers more time resources and degree of freedom for the UAV to adjust its trajectory for maximizing the average system throughput. In Fig. \ref{Fig:6b}, we can observe that the average system throughput decreases with the increase of the grid number for $T=50$ s, which is due to the fact that the mobility of the UAV is affected by the maximum flying speed. In our proposed designs, the UAV moves along the geometric center of the grid, and the maximum flying speed depends on the maximum distance between adjacent geometric centers. The greater the maximum distance, the greater the maximum flying speed. Given a ground area, increasing the grid number means a decrease in the maximum distance between adjacent geometric centers, thereby reducing the maximum flying speed of the UAV. A lower maximum flying speed will make the UAV slower to reach the optimal locations that maximize the average system throughput, thus decreasing the average system throughput.

In Fig. \ref{Fig:6}, compared to the FixST scheme, both the OptT1 and OptT2 schemes achieve higher average system throughput under different period and grid number settings, which demonstrates that the trajectory optimization plays a very important role in improving the average system throughput. The OptT1 scheme can achieve higher average system throughput compared to the OptT2 and FixST schemes, and the performance gap between the OptT1 scheme and the other two schemes increases with the increase of period or the decrease of grid number, which indicates that the advantages of the proposed OptT1 scheme become much more evident with the period increasing or with  the grid number decreasing. Besides, we can also observe that the average system throughput when only communication is performed (i.e., $\bm{\psi}=\mathbf{0}$) is higher than that when ISAC is performed (i.e., $\bm{\psi}=\mathbf{1}$). This is because that sensing will compete for the limited radio resources with communication and interfere with communication as well as restrict the freedom of the UAV trajectory adjustment, thus compromising the average system throughput.

\subsection{Performance Comparisons of Different Sensing Policies}
\begin{figure}[!t]
\renewcommand{\figurename}{Fig.}
  \centering
  \includegraphics[width=9cm]{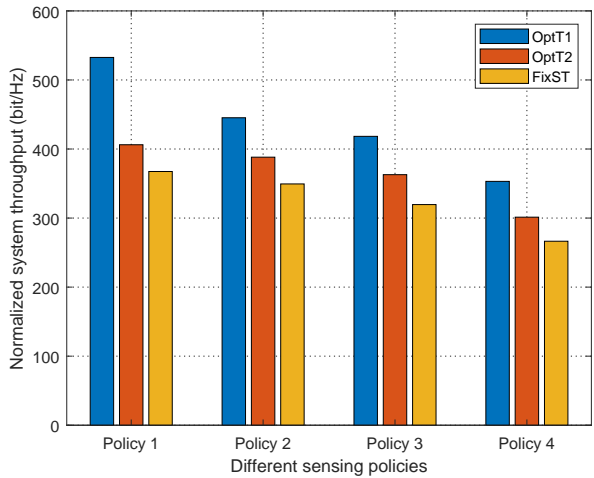}
  \caption{Average system throughput of different schemes versus different sensing policies.}
\label{Fig:7}
\end{figure}

In some practical applications, such as the low-speed target detection or low-precision target tracking, it may be unnecessary to perform sensing all the time. Therefore, sensing should be performed adaptively according to the practical sensing requirements to avoid a waste of resources. Fig. \ref{Fig:7} illustrates the average system throughput under four different sensing policies: Policy 1: No sensing; Policy 2: The sensing interval is variable, which is configured by the proposed AISAC mechanism; Policy 3: The sensing interval is set to fixed two time slots; Policy 4: The sensing interval is set to fixed one time slot. In Policy 2, the sequential sensing intervals within the period $T$ are set to $1$, $2$, $4$, $8$, $\ldots$, $8$ (time slots).

In Fig. \ref{Fig:7}, the OptT1 scheme always has the highest performance under different policies compared with the other two schemes. It is observed that the Policy 1 has the highest average system throughput. This is because communication alone occupies available radio resources and does not suffer interference from the sensing signals. In contrast, the average system throughput of the Policy 4 is lowest. In the Policy 4, sensing will compete with communication for the limited radio resources and interfere with communication, as well as restrict the UAV trajectory adjustment, thus resulting in a sharp decrease of the average system throughput. Compared with the Policy 3, the Policy 2 can significantly improve the average system throughput by allowing high flexibility in the sensing interval configuration. The sensing interval of the Policy 3 and Policy 4 is set as fixed two time slots and one time slot, respectively. By comparison, it can be seen that the larger the sensing interval, the higher the average system throughput. Besides, from the Policy 1 to the Policy 4, the sensing duration increases gradually. The longer sensing duration means consuming more radio resources and also introduces long-term interference to the communication, thus resulting in a considerable decrease in the average system throughput. The above results and analysis demonstrate that the proposed AISAC mechanism can prominently improve the average system throughput by flexibly configuring the sensing interval.

\subsection{Performance Comparisons of Different Antenna Numbers }
In the following, we compare the average system throughput of different optimization schemes under different antenna numbers. Compared with the OptT1 scheme, in the OptT1-equal power scheme, the equal power allocation instead of the optimized power is adopted. Similar definitions are for the OptT2-equal power and FixST-equal power schemes. In the OptT1-random beamforming scheme, the random beamforming is adopted instead of the optimized beamforming. The OptT2-random beamforming and FixST-random beamforming schemes have similar definitions to the OptT1-random beamforming scheme.

\begin{figure}
\renewcommand{\figurename}{Fig.}
\centering
\subfigure[$\bm{\psi}=\mathbf{0}$.]{
\begin{minipage}{7.8cm}
\includegraphics[width=1\textwidth]{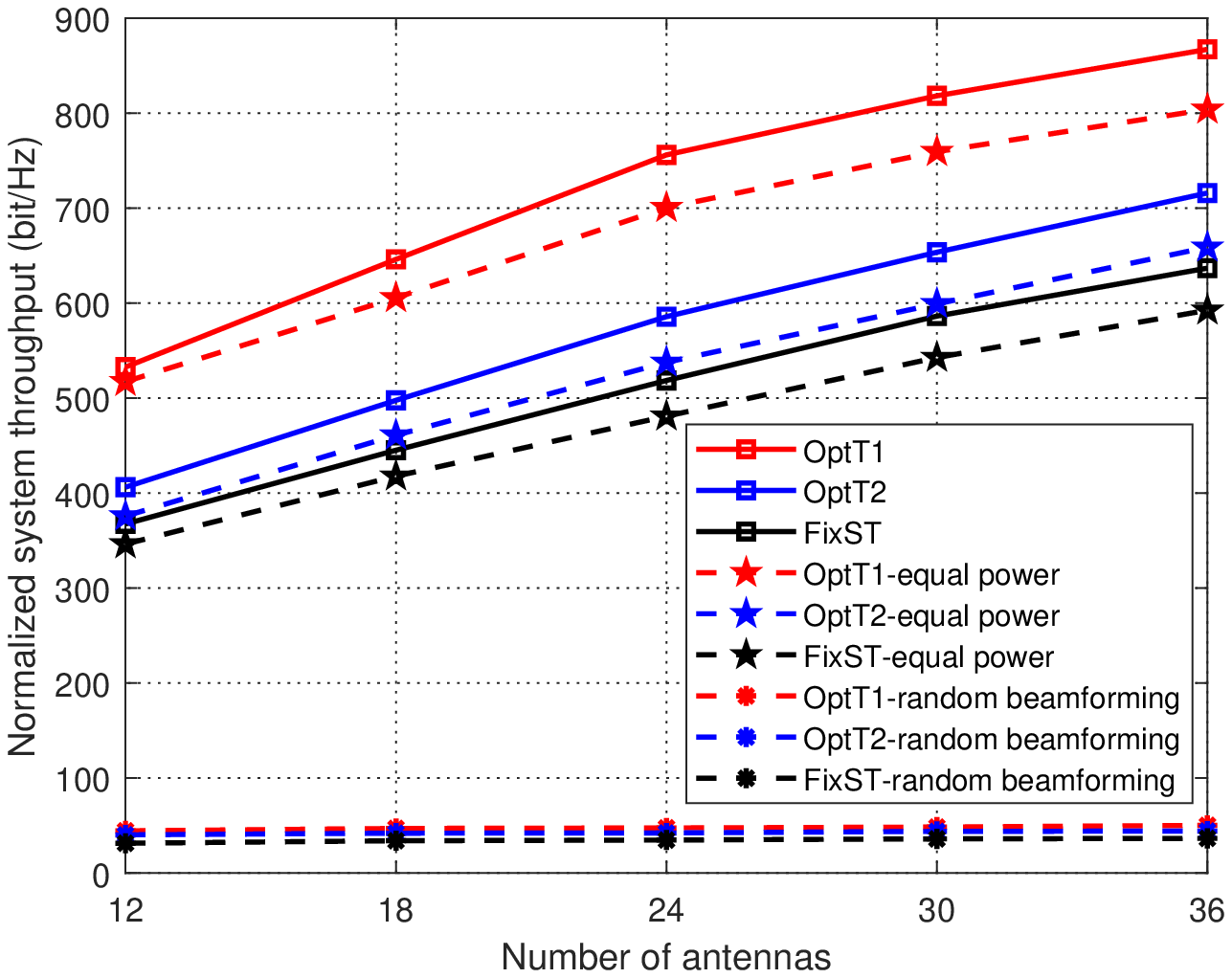}
\label{Fig:8a}
\end{minipage}
}
\subfigure[$\bm{\psi}=\mathbf{1}$.]{
\begin{minipage}{7.8cm}
\includegraphics[width=1\textwidth]{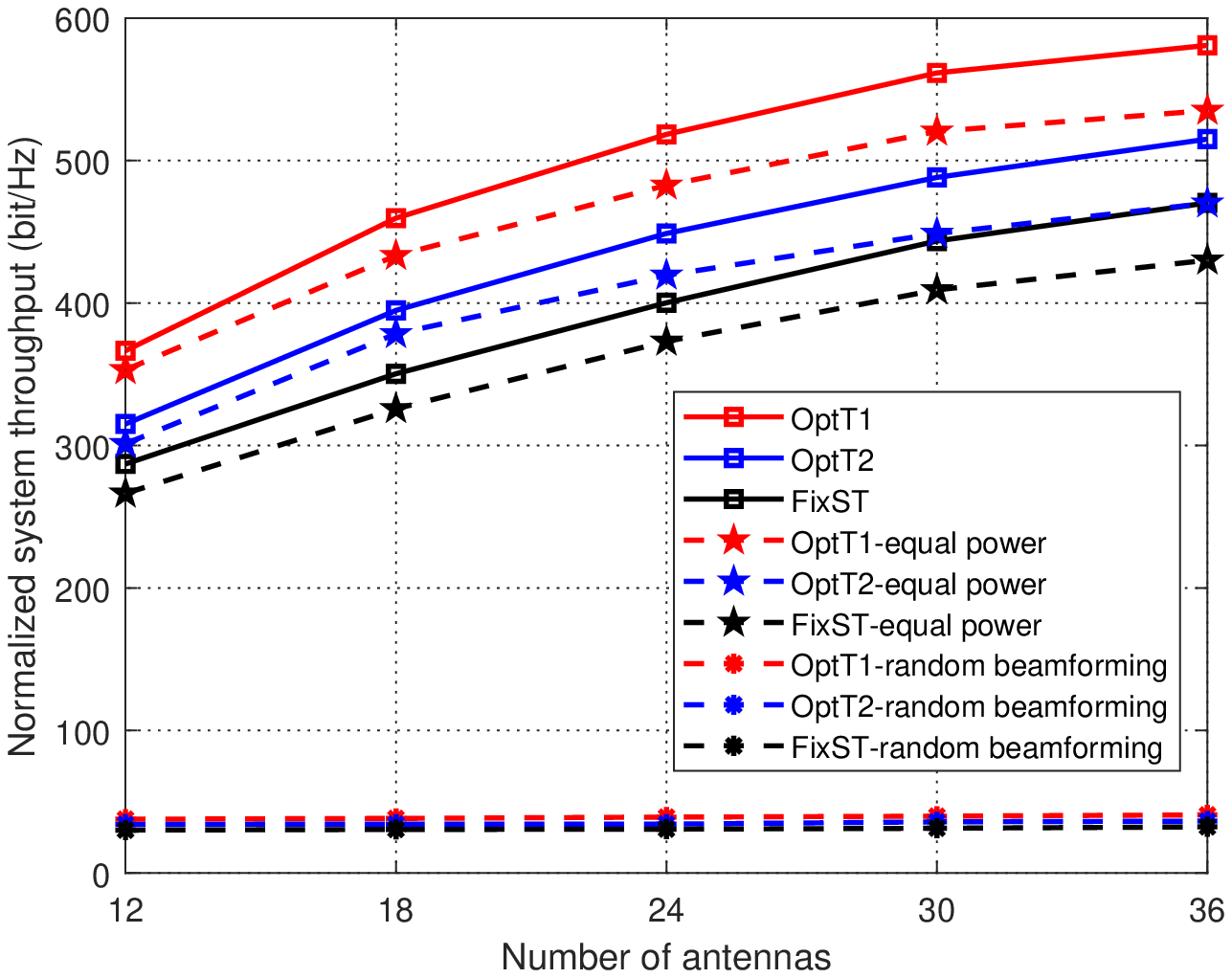}
\label{Fig:8b}
\end{minipage}
}
\caption{(a) Average system throughput of different schemes versus different antenna numbers when $\bm{\psi}=\mathbf{0}$. (b) Average system throughput of different schemes versus different antenna numbers when $\bm{\psi}=\mathbf{1}$.}
\label{Fig:8}
\end{figure}

In both Fig. \ref{Fig:8a} and Fig. \ref{Fig:8b}, the OptT1-equal power, OptT2-equal power, and FixST-equal power schemes significantly outperform the OptT1-random beamforming, OptT2-random beamforming, and FixST-random beamforming schemes, respectively. The average system throughput gap between the OptT1-equal power and OptT1-random beamforming schemes increases as the number of antennas increases. Similar performance characteristics also appear between the OptT2-equal power and OptT2-random beamforming schemes, and between the FixST-equal power and FixST-random beamforming schemes. The reason is that more antennas with the beamforming optimization can offer higher beamforming gain. The effects of power optimization are also studied in Fig. \ref{Fig:8a} and Fig. \ref{Fig:8b}. The average system throughput achieved by the OptT1, OptT2, and FixST schemes is higher than that achieved by the OptT1-equal power, OptT2-equal power, and FixST-equal power schemes, respectively, which indicates that the power optimization plays an important role in improving the average system throughput. In addition, one can also find that for any scheme, the average system throughput when $\bm{\psi}=\mathbf{0}$ is always higher than that when $\bm{\psi}=\mathbf{1}$, which is consistent with the conclusion in the previous subsection.
\section{Conclusion}\label{section:5}
In this work, we investigated the system design and performance optimization for the UAV-enabled ISAC system. First, we proposed a novel AISAC mechanism, in which the sensing duration can be flexibly configured according to the specific sensing requirements instead of maintaining a fixed configuration. Then, a joint beamforming and UAV trajectory optimization problem was formulated to maximize the average system throughput, subject to the QoS requirements of communication and sensing, and UAV's transmit power and location constraints. Furthermore, we proposed a low-complexity optimization algorithm for the considered problem, which sequentially searches the optimal UAV location until reaching the final location of the UAV. Numerical results showed that our proposed designs significantly outperform other benchmark schemes and also confirmed the superiority of the proposed AISAC mechanism. How to extend our designs to other scenarios, e.g., multi-UAV ISAC scenarios, and achieve real-time sensing duration configuration and resource allocation, e.g., using artificial intelligence-based approaches, are worthwhile future works.

\bibliography{IEEEabrv,reference}

\begin{thebibliography}{10}
\providecommand{\url}[1]{#1}
\csname url@samestyle\endcsname
\providecommand{\newblock}{\relax}
\providecommand{\bibinfo}[2]{#2}
\providecommand{\BIBentrySTDinterwordspacing}{\spaceskip=0pt\relax}
\providecommand{\BIBentryALTinterwordstretchfactor}{4}
\providecommand{\BIBentryALTinterwordspacing}{\spaceskip=\fontdimen2\font plus
\BIBentryALTinterwordstretchfactor\fontdimen3\font minus
  \fontdimen4\font\relax}
\providecommand{\BIBforeignlanguage}[2]{{%
\expandafter\ifx\csname l@#1\endcsname\relax
\typeout{** WARNING: IEEEtran.bst: No hyphenation pattern has been}%
\typeout{** loaded for the language `#1'. Using the pattern for}%
\typeout{** the default language instead.}%
\else
\language=\csname l@#1\endcsname
\fi
#2}}
\providecommand{\BIBdecl}{\relax}
\BIBdecl

\bibitem{8972666Survey}
Z.~Feng, Z.~Fang, Z.~Wei, X.~Chen, Z.~Quan, and D.~Ji, ``Joint radar and
  communication: A survey,'' \emph{China Commun.}, vol.~17, no.~1, pp. 1--27,
  Jan. 2020.

\bibitem{8999605_MIMO}
F.~Liu, C.~Masouros, A.~P. Petropulu, H.~Griffiths, and L.~Hanzo, ``Joint radar
  and communication design: Applications, state-of-the-art, and the road
  ahead,'' \emph{{IEEE} Trans. Commun.}, vol.~68, no.~6, pp. 3834--3862, Jun.
  2020.

\bibitem{9585321Survey}
J.~A. Zhang, M.~L. Rahman, K.~Wu, X.~Huang, Y.~J. Guo, S.~Chen, and J.~Yuan,
  ``Enabling joint communication and radar sensing in mobile networks-{A}
  survey,'' \emph{IEEE Commun. Surveys Tuts.}, vol.~24, no.~1, pp. 306--345,
  1st Quart. 2022.

\bibitem{9124713_MIMO}
X.~Liu, T.~Huang, N.~Shlezinger, Y.~Liu, J.~Zhou, and Y.~C. Eldar, ``Joint
  transmit beamforming for multiuser {MIMO} communications and {MIMO} radar,''
  \emph{{IEEE} Trans. Signal Process.}, vol.~68, pp. 3929--3944, Jun. 2020.

\bibitem{9724174GeneralizedBeamforming}
L.~Chen, Z.~Wang, Y.~Du, Y.~Chen, and F.~R. Yu, ``Generalized transceiver
  beamforming for {DFRC} with {MIMO} radar and {MU-MIMO} communication,''
  \emph{{IEEE} J. Sel. Areas Commun.}, vol.~40, no.~6, pp. 1795--1808, Jun.
  2022.

\bibitem{8288677_MIMO}
F.~Liu, C.~Masouros, A.~Li, H.~Sun, and L.~Hanzo, ``{MU-MIMO} communications
  with {MIMO} radar: From co-existence to joint transmission,'' \emph{{IEEE}
  Trans. Wireless Commun.}, vol.~17, no.~4, pp. 2755--2770, Apr. 2018.

\bibitem{horizontal_locations}
\BIBentryALTinterwordspacing
H.~Hua, J.~Xu, and T.~X. Han, ``Optimal transmit beamforming for integrated
  sensing and communication,'' \emph{arXiv:2104.11871}, 2021. [Online].
  Available: \url{https://arxiv.org/abs/2104.11871}
\BIBentrySTDinterwordspacing

\bibitem{fang2022joint}
\BIBentryALTinterwordspacing
X.~Fang, W.~Feng, Y.~Chen, N.~Ge, and Y.~Zhang, ``Joint communication and
  sensing: Models and potentials of using {MIMO},'' \emph{arXiv:2205.09409},
  2022. [Online]. Available: \url{https://arxiv.org/abs/2205.09409}
\BIBentrySTDinterwordspacing

\bibitem{9416177RIS}
X.~Wang, Z.~Fei, Z.~Zheng, and J.~Guo, ``Joint waveform design and passive
  beamforming for {RIS}-assisted dual-functional radar-communication system,''
  \emph{{IEEE} Trans. Veh. Technol.}, vol.~70, no.~5, pp. 5131--5136, May 2021.

\bibitem{9769997RIS}
R.~Liu, M.~Li, Y.~Liu, Q.~Wu, and Q.~Liu, ``Joint transmit waveform and passive
  beamforming design for {RIS}-aided {DFRC} systems,'' \emph{{IEEE} J. Sel.
  Topics Signal Process.}, pp. 1--1, May 2022.

\bibitem{9591331RIS}
X.~Wang, Z.~Fei, J.~Huang, and H.~Yu, ``Joint waveform and discrete phase shift
  design for {RIS}-assisted integrated sensing and communication system under
  {Cramer-Rao Bound} constraint,'' \emph{{IEEE} Trans. Veh. Technol.}, vol.~71,
  no.~1, pp. 1004--1009, Jan. 2022.

\bibitem{9449980AI}
S.~Huang, M.~Zhang, Y.~Gao, and Z.~Feng, ``{MIMO} radar aided mmwave
  time-varying channel estimation in {MU-MIMO} {V2X} communications,''
  \emph{{IEEE} Trans. Wireless Commun.}, vol.~20, no.~11, pp. 7581--7594, Nov.
  2021.

\bibitem{9492131AI}
J.~Mu, Y.~Gong, F.~Zhang, Y.~Cui, F.~Zheng, and X.~Jing, ``Integrated sensing
  and communication-enabled predictive beamforming with deep learning in
  vehicular networks,'' \emph{{IEEE} Commun. Lett.}, vol.~25, no.~10, pp.
  3301--3304, Oct. 2021.

\bibitem{7888557_EnergyEfficientUAV}
Y.~Zeng and R.~Zhang, ``Energy-efficient {UAV} communication with trajectory
  optimization,'' \emph{{IEEE} Trans. Wireless Commun.}, vol.~16, no.~6, pp.
  3747--3760, Jun. 2017.

\bibitem{7572068RelayUAV}
Y.~Zeng, R.~Zhang, and T.~J. Lim, ``Throughput maximization for {UAV}-enabled
  mobile relaying systems,'' \emph{{IEEE} Trans. Commun.}, vol.~64, no.~12, pp.
  4983--4996, Dec. 2016.

\bibitem{8618602SecuringUAV}
G.~Zhang, Q.~Wu, M.~Cui, and R.~Zhang, ``Securing {UAV} communications via
  joint trajectory and power control,'' \emph{{IEEE} Trans. Wireless Commun.},
  vol.~18, no.~2, pp. 1376--1389, Feb. 2019.

\bibitem{8247211MUltiUAV}
Q.~Wu, Y.~Zeng, and R.~Zhang, ``Joint trajectory and communication design for
  multi-{UAV} enabled wireless networks,'' \emph{{IEEE} Trans. Wireless
  Commun.}, vol.~17, no.~3, pp. 2109--2121, Mar. 2018.

\bibitem{8647530UAVSensing}
H.~Sallouha, M.~M. Azari, and S.~Pollin, ``Energy-constrained {UAV} trajectory
  design for ground node localization,'' in \emph{Proc. IEEE Global Commun.
  Conf. (GLOBECOM)}, Dec. 2018, pp. 1--7.

\bibitem{9409835UAVSensing}
I.~Bisio, C.~Garibotto, H.~Haleem, F.~Lavagetto, and A.~Sciarrone, ``On the
  localization of wireless targets: A drone surveillance perspective,''
  \emph{{IEEE} Netw.}, vol.~35, no.~5, pp. 249--255, Sep. 2021.

\bibitem{8894454UAVSensing}
Z.~Wang, R.~Liu, Q.~Liu, J.~S. Thompson, and M.~Kadoch, ``Energy-efficient data
  collection and device positioning in {UAV}-assisted {IoT},'' \emph{IEEE
  Internet Things J.}, vol.~7, no.~2, pp. 1122--1139, Feb. 2020.

\bibitem{lyu2021joint}
\BIBentryALTinterwordspacing
Z.~Lyu, G.~Zhu, and J.~Xu, ``Joint maneuver and beamforming design for
  {UAV}-enabled integrated sensing and communication,''
  \emph{arXiv:2110.02857}, 2021. [Online]. Available:
  \url{https://arxiv.org/abs/2110.02857}
\BIBentrySTDinterwordspacing

\bibitem{9293257Constrained}
X.~Wang, Z.~Fei, J.~A. Zhang, J.~Huang, and J.~Yuan, ``Constrained utility
  maximization in dual-functional radar-communication multi-{UAV} networks,''
  \emph{{IEEE} Trans. Commun.}, vol.~69, no.~4, pp. 2660--2672, Apr. 2021.

\bibitem{skolnik2008radar}
M.~I. Skolnik, \emph{Radar {Handbook}}.\hskip 1em plus 0.5em minus 0.4em\relax
  McGraw-Hill Education, 2008.

\bibitem{9557830YuanWeijie_twoway}
W.~Yuan, Z.~Wei, S.~Li, J.~Yuan, and D.~W.~K. Ng, ``Integrated sensing and
  communication-assisted orthogonal time frequency space transmission for
  vehicular networks,'' \emph{{IEEE} J. Sel. Topics Signal Process.}, vol.~15,
  no.~6, pp. 1515--1528, Nov. 2021.

\bibitem{9171304LiuFan_twoway}
F.~Liu, W.~Yuan, C.~Masouros, and J.~Yuan, ``Radar-assisted predictive
  beamforming for vehicular links: Communication served by sensing,''
  \emph{{IEEE} Trans. Wireless Commun.}, vol.~19, no.~11, pp. 7704--7719, Nov.
  2020.

\bibitem{9226446Gain}
M.~Temiz, E.~Alsusa, and M.~W. Baidas, ``A dual-functional massive {MIMO}
  {OFDM} communication and radar transmitter architecture,'' \emph{{IEEE}
  Trans. Veh. Technol.}, vol.~69, no.~12, pp. 14\,974--14\,988, Dec. 2020.

\bibitem{Fractional_Programming}
K.~Shen and W.~Yu, ``Fractional programming for communication systems-{Part I}:
  Power control and beamforming,'' \emph{{IEEE} Trans. Signal Process.},
  vol.~66, no.~10, pp. 2616--2630, May 2018.

\bibitem{Semidefinite_Relaxation}
Z.-q. Luo, W.-k. Ma, A.~M.-c. So, Y.~Ye, and S.~Zhang, ``Semidefinite
  relaxation of quadratic optimization problems,'' \emph{{IEEE} Signal Process.
  Mag.}, vol.~27, no.~3, pp. 20--34, May 2010.

\bibitem{9481926Rank}
X.~Xu, Y.-C. Liang, G.~Yang, and L.~Zhao, ``Reconfigurable intelligent surface
  empowered symbiotic radio over broadcasting signals,'' \emph{{IEEE} Trans.
  Commun.}, vol.~69, no.~10, pp. 7003--7016, Oct. 2021.

\bibitem{9570143Rank}
X.~Mu, Y.~Liu, L.~Guo, J.~Lin, and R.~Schober, ``Simultaneously transmitting
  and reflecting {(STAR)} {RIS} aided wireless communications,'' \emph{{IEEE}
  Trans. Wireless Commun.}, vol.~21, no.~5, pp. 3083--3098, May 2022.

\bibitem{upper_bound_approximation}
J.~Zuo, Y.~Liu, Z.~Qin, and N.~Al-Dhahir, ``Resource allocation in intelligent
  reflecting surface assisted {NOMA} systems,'' \emph{{IEEE} Trans. Commun.},
  vol.~68, no.~11, pp. 7170--7183, Nov. 2020.

\bibitem{boyd2004convex}
S.~Boyd, S.~P. Boyd, and L.~Vandenberghe, \emph{Convex {Optimization}}.\hskip
  1em plus 0.5em minus 0.4em\relax Cambridge, U.K.: Cambridge {Univ.} {Press},
  2004.

\end{thebibliography}
\bibliographystyle{IEEEtran}

\end{document}